\newdimen\captionwidth 
\long\def\@makecaption#1#2{ 
\vskip 10pt 
\captionwidth=14cm 
\begin{center} 
\begin{minipage}[t]{\captionwidth}
{\footnotesize 
 #1:\;#2}
\end{minipage}
\end{center}} 
\documentclass[12pt]{article}
\usepackage{epsfig,amsmath,amssymb,graphics,color,calc,cite} 
\usepackage{amsmath} 
\usepackage{amsthm}
\usepackage{amssymb}
\usepackage{amsopn}
\newcommand{\sezione}[2]{ 
\refstepcounter{section}\label{#2} 
\setcounter{equation}{0} 
\setcounter{subsection}{0} 
\addcontentsline{toc}{section} 
      {\normalsize\textbf{\thesection.\ #1}} 
\bigskip\bigskip\noindent 
\normalsize\textbf{\thesection.\ #1}\nopagebreak\smallskip\nopagebreak} 
\def\thesection{{\normalsize\arabic{section}}} 
\newcounter{appendice}

\newtheorem{teo}{Theorem}[section]	\newtheorem{pro}[teo]{Proposition}
\newtheorem{defi}[teo]{Definition}	\newtheorem{lem}[teo]{Lemma}
\newtheorem{cor}[teo]{Corollary}	\newtheorem{rem}[teo]{Remark}
\newtheorem{con}[teo]{Condition}

\newcommand{\bteo}[1]{\begin{teo}\label{#1}}
\newcommand{\bpro}[1]{\begin{pro}\label{#1}}
\newcommand{\bdefi}[1]{\begin{defi}\label{#1}}
\newcommand{\blem}[1]{\begin{lem}\label{#1}}
\newcommand{\bcor}[1]{\begin{cor}\label{#1}}
\newcommand{\brem}[1]{\begin{rem}\label{#1}}
\newcommand{\bcon}[1]{\begin{con}\label{#1}}

\newcommand{\eteo}{\end{teo}}	\newcommand{\epro}{\end{pro}}
\newcommand{\edefi}{\end{defi}}	\newcommand{\elem}{\end{lem}}
\newcommand{\ecor}{\end{cor}}	\newcommand{\erem}{\end{rem}}
\newcommand{\econ}{\end{con}}

\renewcommand{\qed}{\hfill $\Box$}
\renewcommand{\eqref}[1]{(\ref{#1})}

\newcommand{\be}[1]{\begin{equation}\label{#1}}
\newcommand{\bea}[1]{\begin{eqnarray}\label{#1}}
\newcommand{\besn}{\begin{equation*}}
\newcommand{\beasn}{\begin{eqnarray*}}

\newcommand{\su}{\subset}	
\newcommand{\sm}{\setminus}	\newcommand{\es}{\emptyset}

\newcommand{\dis}{\mathop{\rm d}\nolimits}

\newcommand{\gap}{\mathop{\rm gap}\nolimits}

\newcommand{\na}{\nabla}

	\newcommand{\id}{{1 \mskip -5mu {\rm I}}}
 
\newcommand{\noi}{\noindent}

\newcommand{\ol}[1]{\overline{#1}}
		\renewcommand{\d}{\delta}	
\newcommand{\e}{\varepsilon}	\newcommand{\f}{\varphi} 
\newcommand{\g}{\gamma}		\newcommand{\h}{\eta}

	\renewcommand{\r}{\rho}
		
		\renewcommand{\th}{\vartheta}
	
\newcommand{\z}{\zeta}

		\renewcommand{\L}{\Lambda}
	\renewcommand{\O}{\Omega}

\newcommand{\cA}{\mathcal A}	\newcommand{\cB}{\mathcal B} 
	 
\newcommand{\cE}{\mathcal E}

	\newcommand{\cL}{\mathcal L}

\newcommand{\cQ}{\mathcal Q}

\newcommand{\bE}{\mathbb E}

	\newcommand{\bR}{\mathbb R}

	\newcommand{\bZ}{\mathbb Z} 
\newcommand{\newatop}[2]{\genfrac{}{}{0pt}{}{#1}{#2}} 
\renewcommand{\complement}{{\textrm{c}}}

\newcommand{\TR}{T^{{R}}}
\newcommand{\TL}{T^{{L}}}

\begin{document}

\begin{titlepage}
\par\vskip 1cm\vskip 2em
\begin{center}

{\LARGE Exclusion processes with degenerate rates: 
 $\vphantom{\bigg\{}$\\
convergence to equilibrium and tagged particle*
\par
}
%\vskip 2.5em 

\vskip 3.0em 

\lineskip .5em
{\large
\begin{tabular}[t]{c}
$\mbox{Lorenzo Bertini}^{1} \phantom{merda} \mbox{Cristina Toninelli}^{2}$
\\
\end{tabular}
\par
}

\medskip
{\small
\begin{tabular}[t]{ll}
{\bf 1} & {\it 
Dipartimento di Matematica, Universit\`a di Roma La Sapienza}\\
&  P.le A.\ Moro 2, 00185 Roma, Italy\\
&  E--mail: {\tt bertini@mat.uniroma1.it}\\
\\
{\bf 2} & {\it Dipartimento di Fisica, Universit\`a di Roma La Sapienza}\\
&  P.le A.\ Moro 2, 00185 Roma, Italy\\
&  E--mail: {\tt cristina.toninelli@roma1.infn.it}
\end{tabular}
}

\bigskip
%\vskip .5em { September 2000}
\end{center}

\vskip 1 em

\centerline{\bf Abstract} 
\smallskip
Stochastic lattice gases with degenerate rates, namely conservative
particle systems where the exchange rates vanish for some
configurations, have been introduced as simplified models for glassy
dynamics.
We introduce two particular models and consider them in a finite
volume of size $\ell$ in contact with particle reservoirs at the
boundary.
We prove that, as for non--degenerate rates, the inverse of the
spectral gap and the logarithmic Sobolev constant grow as $\ell^2$.
It is also shown how one can obtain, via a scaling limit from the
logarithmic Sobolev inequality, the exponential decay of a macroscopic entropy
associated to a degenerate parabolic differential equation (porous media
equation). 
We analyze finally the tagged particle displacement for the stationary
process in infinite volume.  In dimension larger than two we prove
that, in the diffusive scaling limit, it converges to a Brownian motion
with non--degenerate diffusion coefficient.

%\noindent{\it Running title: }
\vskip 0.8 em

\vfill
\noindent    
{\bf MSC2000:} 60K35, 82C20, 60F17, 35K65. 
% {\bf PACS2003} 61.43.Fs, 02.50.Ga, 05.50.+q.
\vskip 0.8 em
\noindent
{\bf Keywords:}\ 
Exclusion processes, Spectral gap, Logarithmic Sobolev inequalities, 
Porous media equation, Tagged particle diffusion.

\bigskip\bigskip
\footnoterule
\vskip 1.0em
{\small 
\noindent
We acknowledge the support of Cofinanziamento MURST.

\smallskip
\noindent
$*$)
Dedicated to Gianni Jona-Lasinio, whom 
{\em debeo, quod in cogitationem veni, mihi morum emendatione et
curatione opus esse; quod neque ad aemulationem sophisticam declinavi,
neque de theorematis commentatus sum...\/} [Marcus Aurelius]

\vskip 1.0em
}
\end{titlepage}
\eject

\sezione{Introduction}{s:intro}
\par\noindent
We analyze some models of stochastic lattice gases with hard core
exclusion, i.e.\ systems of particles on a lattice $\Lambda\su \bZ^d$
with the constraint that on each site there is at most one particle.
A configuration is therefore defined by giving for each site
$x\in\Lambda$ the occupation number $\eta_x\in \{0,1\}$, which
represents an empty or occupied site.
The dynamics is given by a continuous time Markov process, which allows 
the exchange of the occupation numbers across a bond $\{x,y\}$  of
neighboring sites $x$ and $y$ with a rate $c_{x,y}(\h)$ depending on
the configuration $\h$.   
The simplest case is the symmetric simple exclusion process
(SEP), in which $c_{x,y}(\h)=1$.  We instead consider processes
in which there are some constraints in order for the exchange to be
allowed, i.e.\ the rate $c_{x,y}(\h)$ degenerates for some configurations
$\h$. We mention that Glauber--like, non--conservative, models with
degenerate rates have been studied in \cite{AD}.
We shall discuss conservative stochastic lattice gases with degenerate
rates both in infinite volume, $\L=\bZ^d$, and on a bounded domain. In
the latter case we shall however allow exchanges of particles with
external reservoirs at the boundary.

This kind of interacting particle systems, called {\em kinetically
constrained lattice gases,\/} have been introduced in the
physical literature as simplified models for some peculiar phenomena of the
``glassy'' dynamics \cite{ReviewKLG}. Let us recall the physical problem
\cite{DeBenedetti}. 
Experimentally, a glass can be obtained by cooling a liquid fast
enough in order to avoid crystallization. Below the melting
temperature the liquid enters a metastable phase in which the
relaxation time is rather long and increase dramatically if the
temperature is further lowered.  When the time to reach equilibrium
becomes longer than the experimentally accessible time scales, the
liquid freezes in an amorphous solid phase, which is called glass.
A complete theoretical explanation of this ``glass transition'' is still
lacking \cite{Mezard}.  
The first question to be settled is whether the glass is a new state
of matter or a long lived metastable state, i.e.\ whether 
the dramatic increase of the relaxation time is due to an underlying
equilibrium transition or is a dynamical phenomenon.
Since no static divergent correlation length is detected and the
structural properties show a very small temperature dependence
\cite{DeBenedetti},  
it is possible that a purely dynamical transition takes place. 
Thus a major goal is to understand the mechanism inducing the 
{\em dynamical arrest\/} which prevents the relaxation of the system. 
This should be related to other phenomena such as the stretched exponential decay of the structure
function for temperatures close to the glass transition and the aging
phenomena for supercooled liquids quenched to lower temperatures
\cite{Struick,Leticia}.

In this context, several approaches have been proposed.
We mention the {\em random first order scenario\/}
\cite{KT,KTW} and the {\em mode coupling theory\/} \cite{Gotze}; 
from them the following picture in the mean field approximation arises
\cite{Leticia}.  
There exists a finite temperature $T_d$ such that for
$T>T_d$ the configuration space consists of a single ergodic
component, while for $T<T_d$ it is broken into many disconnected
ergodic components. Moreover the relaxation time diverges as 
$T\downarrow T_d$.
An open question is how to go beyond this mean field picture, 
in particular to establish whether this ergodic/non--ergodic transition
is an effect of the mean field approximation or is related to
the behavior of real glasses. 
The kinetically constrained particle systems, originally introduced to
get some insight on the physical mechanism for the dynamical arrest
\cite{KA}, have recently \cite{Barrat,TBF} also been used to
investigate how the mean field scenario has to be modified for short
range models.

The basic idea behind these models is that the motion of a molecule is
inhibited by geometrical constraints due to the presence of
surrounding particles.  In particular, a molecule can be caged by
neighboring ones and the cage must be opened to allow its motion.  It
is thus possible that these local constraints might produce, for a
finite value of the density, a cooperative behavior inducing the
slowing down of the dynamics.  The kinetically constrained lattice
gases, are therefore defined  by choosing exchange rates $c_{x,y}$ which
 encode this cage effect.
Despite their simplicity and the discrete character, they might
capture, at least at a mesoscopic level, some of the key dynamical
ingredients of real glasses.  Moreover, they have recently been used
\cite{sellitto} to model granular systems, which is another class of
systems displaying a glass--like dynamical arrest, known as jamming
transition \cite{Jaeger, Liu}.

One of the most studied among such models is the so called Kob--Andersen
model (KA) \cite{KA}.  
Let $m=1,\dots,2d-1$, the KA model is defined by choosing
$c_{x,y}(\h)=1$ if at least $m$ nearest neighbors of $x$ different
from $y$ are empty and $m$ nearest neighbors of $x$ different from $y$
are also empty, $c_{x,y}(\h)=0$ otherwise.  
Note that for $m=0$ we would recover the symmetric simple exclusion.
Let us consider this model for $d\ge 2$ in a finite cube $\L\su\bZ^d$
of size $\ell$; it satisfies the detailed balance w.r.t.\ the uniform
measure on the hyperplanes with fixed total number of
particles. However it is not ergodic, for instance if $d=2$ and $m=1$
two fully occupied consecutive rows do not move.
By the general theory of finite state Markov chain it is however
possible to decompose the state space into irreducible components.  
A natural question is the asymptotic behavior, as $\ell\to\infty$, of
the probability $p_\ell(\rho)$ of the ``maximal irreducible
component'', here $\r$ is the density of particles on the hyperplane.
The ergodic/non--ergodic transition at a density $\r_c\in [0,1]$ would
correspond to $p_\infty(\rho)=1$ for $\r<\r_c$ and $p_\infty(\rho)<1$
for $\r>\r_c$.  
An alternative definition is obtained by looking at the model directly
in infinite volume.  
We note the process satisfies detailed balance
w.r.t.\ each Bernoulli product measure $\mu_\r$, $\r\in [0,1]$ but
there are other invariant measures, for instance some 
concentrated on single configurations. 
Denoting by $P_t$ the semigroup associated to the process,
by the spectral theorem, we have that 
$$ 
\lim_{t\to\infty} \: \int\!d\mu_\r(\h) \: 
\big[ P_t f(\h) - \mu_\r(f) \big]^2  =0 
\quad {\textrm{ for any }} f\in L_2(\mu_\r) 
$$ 
if and only if zero is a 
simple eigenvalue of the generator; 
in such a case we say that the process is ergodic in $L_2(\mu_\r)$.
If the process is ergodic for $\r<\widehat\r_c$
and not ergodic for $\r>\widehat\r_c$ we would then say that the
ergodic/non--ergodic transition occurs at $\widehat\r_c$.

In the case $d=3$ and $m=2$, numerical simulations of the KA model \cite{KA}
suggested that such a transition, in the sense of the former
definition, takes place at $\rho_c\simeq 0.881$.
However in \cite{TBF} it is shown that for $d=2$, $m=1$ and $d=3$,
$m=1,2$ we have $\rho_c=1$ whereas for $d=2$, $m=2,3$ and $d=3$,
$m=3,4,5$ we have $\rho_c=0$.
The finite size corrections are also discussed in
\cite{TBF}; in particular it shown that for $\r$ close to one
the thermodynamic limit is achieved for sizes of order 
$L(\r)= \exp\{c(1-\r)^{-1}\} $ for $d=2$, $m=1$ and  $d=3$, $m=1$ 
whereas $L(\r)=\exp\{\exp\{ c(1-\r)^{-1} \}\}$ for $d=3$, $m=2$.

It is also possible to consider the KA model in a finite volume of
size $\ell$ in contact with particle reservoirs at the boundary,
see \cite{Barrat,Kurchan} for numerical simulations.
The total number of particles is not anymore conserved and, for $m\le
d-1$, the system is ergodic in the whole configuration space and
reversible w.r.t.\ the Bernoulli measure whose density is fixed by the
reservoirs.  However, the dynamical arrest is not ruled out since, as
$\ell\to\infty$, the speed of convergence toward the unique invariant
measure might exhibit a crossover as a function of the density.  A
preliminary question is then the asymptotic behavior of the relaxation
time, which might be defined as the inverse of the spectral gap of the
generator, for $\ell\to\infty$. Note that for SEP the relaxation time
grows as $\ell^2$ uniformly in the density.

Another physically relevant issue is the asymptotic displacement
$x(t)$ of a tagged particle. Indeed, for supercooled liquids near the
glass transition, a decrease of the mean square displacement 
$\bE\big( x(t)^2\big)$ is experimentally detected \cite{DeBenedetti}
with the possibility of a crossover from a diffusive, to a
sub--diffusive 
behavior related to the dynamical arrest.  Let us denote by
$D_{\textrm{self}} = D_{\textrm{self}}(\rho)$ the diffusion
coefficient of a tagged particle.  For the KA model with $d=3$ and
$m=2$ the numerical simulation in \cite{KA} for $\rho<\rho_c\simeq
0.881$ suggested the power law behavior $D_{\textrm{self}} =
(\rho_c-\rho)^\alpha$, with $\alpha\simeq 0.3$.  This evidence is
however probably due to finite size effects; in fact in \cite{TBF} it
is argued that for each $\rho\in [0,1)$ the diffusion coefficient is
strictly positive.

The main results of this paper are sharp asymptotics on the relaxation
time and the diffusive behavior of the tagged particle for some
kinetically constrained lattice gases.  We are not able to prove these
results for the KA model, but they will be obtained for two simpler
models defined by different geometrical constraints.  The first model
is defined by an exchange rate $c_{x,y}$ which vanishes if the two
neighbors of the bond $\{x,y\}$ along the direction $y-x$ are both
occupied. In the second model the exchange rate $c_{x,y}$ across the
bond $\{x,y\}$ vanishes if more than half of its neighboring sites,
i.e.\ more than $2d-1$ neighbors, are occupied.  These models are in
the same spirit of KA, but the degeneracy of the rates is not
comparable. Indeed, there are exchanges allowed for KA and forbidden
for our models as well as the converse. There is however an important
simplifying feature of the models defined above, which plays an
essential role in the rigorous analysis: it is possible to construct a
finite cluster of empty sites which, uniformly in the configuration on
its complement, can be shifted using only allowed exchanges.  
The results of this paper show that the models above defined behave
essentially as the simple exclusion process and therefore they are not
really appealing as models for the glass transition.
The asymptotic as $\rho\uparrow 1$ of the relaxation time and of
$D_{\textrm{self}}(\rho)$ are however different from SEP.  
Of course, an interesting issue is whether the results obtained in this
paper holds also for the KA model, in other words if the simplifying
feature mentioned above is only a technical need or the behavior of KA
is essentially different. As argued in \cite{TBF}, to which we refer
for a further discussion on this point, it is expected that also KA is
ergodic in $L_2(\mu_\r)$ and the tagged particle diffusion is not
degenerate for each $\r\in [0,1)$. The asymptotic of relaxation times
as $\ell\to\infty$ and the behavior of $D_{\textrm{self}}(\rho)$ as
$\r\uparrow 1$ might however be different; their analysis appears to
be a more difficult task.

Another relevant issue is the macroscopic behavior of the kinetically
constrained lattice gases. For non--degenerate rates, the
hydrodynamical limit \cite{S} states that if the initial condition has
a density profile then, under a diffusive rescaling, at later times we
still have a density profile which can be obtained from the initial
one by solving a parabolic equation. For SEP this is simply the heat
equation.  If the rates degenerate a natural candidate for the
hydrodynamic limit is a parabolic equation of porous media type
degenerating when the density approaches one; an analogous result has
indeed been proven for a model in which the occupation number $\h_x$
is a continuous variable \cite{sep}.  For kinetically constrained
lattice gases there is however a serious obstruction: the hydrodynamic
limit cannot hold for any initial condition admitting a density
profile. Consider for instance the models described above in $d=1$ and
take an initial configuration given by a sequence of two occupied
sites and one empty at the left of the origin and three occupied sites
and one empty at the right. This configuration is invariant for the
microscopic dynamic, however the associated density profile evolves
diffusively for the putative macroscopic evolution. Indeed, the
initial density profile is $2/3$ at the left and $3/4$ at the right,
it is thus bounded away from 1 and therefore is not affected by the
degeneracy.  On the other hand such phenomenon is somehow exceptional
and we expect a hydrodynamic behavior for a suitable large class of
initial conditions.

\medskip
\noindent{\it Outline and summary of results.}

In Section \ref{s:n} we define more precisely the models we analyze
and introduce the basic notation. 

In Sections \ref{s:g} and \ref{s:ls} we consider these models on
finite volume of size $\ell$ with reservoirs allowing particle
exchanges at the boundary.  By our choice of the rates the processes
are ergodic and reversible w.r.t.\ the product Bernoulli measure whose
density is fixed by the reservoirs. We then discuss the rate of
convergence to this unique invariant measure.  

In particular, in Section \ref{s:g}, we show that the spectral gap
shrinks to zero as $\ell^{-2}$ when $\ell\to \infty$. An analogous
result for the simple exclusion process on the hyperplane with fixed
number of particles has been obtained in \cite[\S 8]{Q} and has been
extended to Kawasaki dynamics, under suitable mixing conditions on the
invariant measure, in \cite{CM,LY}.  Our proof is based on a
comparison argument with the Glauber dynamics reversible w.r.t.\ the
same Bernoulli measure and the construction of a suitable path which
enables to move a particle from the boundary to any site by using only
allowed exchanges.
 
In Section \ref{s:ls} we show that the above idea can be used also to
show that the logarithmic Sobolev constant, which controls the
exponential decay of the entropy, grows as $\ell^2$. 
An analogous result for the Kawasaki dynamics, under suitable mixing
conditions on the invariant measure, has been obtained in \cite{CMR,Y}.  
We then prove that, via a scaling limit, one can
obtain the exponential decay of a macroscopic entropy 
associated to the porous media equation.

In Section \ref{s:tp} we consider these processes on infinite
volume, in such a case they are reversible w.r.t.\ the Bernoulli
measure $\mu_\r$ and we show they are ergodic in $L_2(\mu_\r)$ for any
$\r\in [0,1]$. We do not discuss the rate of convergence to
equilibrium in this context, which for the exclusion process is
algebraic with diffusive exponent $d/2$, see \cite{BZ,D,JLQY}.
We analyze the displacement $x(t)$ of a tagged particle
for the stationary process.
We recall that for SEP in $d=1$ we have \cite{arr} $\bE \big( x(t)^2
\big) \approx \sqrt{t}$ while for $d\ge 2$ the displacement $x(t)$
satisfies \cite{KV} a central limit theorem with strictly positive 
variance, i.e.\ $\bE \big( x(t)^2 \big) \approx t$.
By the ergodicity of the process in  $L_2(\mu_\r)$, 
one can repeat the arguments in \cite{KV,S} and
show that, under a diffusive rescaling, $x(t)$ converges to a Brownian
motion with diffusion coefficient
$D_{\textrm{self}}=D_{\textrm{self}}(\r)$ given by a variational
formula.  Of course, if $d=1$ we have that $D_{\textrm{self}}=0$ as
for the simple exclusion. By following the strategy in \cite{Sart,S}
we finally prove that for $d\ge 2$ and each $\r\in [0,1)$ we have
$D_{\textrm{self}}>0$.

\sezione{Definition of the models}{s:n}
\par\noindent
The spatial structure is modeled by the $d$--dimensional cubic lattice
$\bZ^d$ in which we let $e_i$, $i=1,\dots,d$ be the coordinate unit
vectors. We denote by $x,y,z$ the sites
of $\bZ^d$ and by $\dis(x,y):=|x-y|$ the Euclidean distance
between $x$ and $y$.  Given $A,B \su \bZ^d$ we then let, as usual,
$\dis(A,B) = \inf\{ \dis(x,y)\,,\; x\in A, y\in B \}$.

For $A\su\bZ^d$, the configuration space in $A$ is $\O_A:=\{0,1\}^A$.
If $A=\bZ^d$, we drop it from the notation, namely we let 
$\O:=\O_{\bZ^d}$. 
We can regard a configuration $\eta\in\O_A$ as a map from $A$ to
$\{0,1\}$; the value $\eta_x\in\{0,1\}$ is interpreted as the number
of particles of the configuration $\eta$ at the site $x$. If
$\h\in\O_A$ and $B\su A$ we denote by $\h_B$ the restriction of the 
configuration $\h$ to $\O_B$. Let $A,B\su\bZ^d$ be disjoint subsets,
$A\cap B=\es$; given $\h\in\O_A$ and $\xi\in\O_B$ we let
$\h\xi\in\O_{A\cup B}$ be the configuration such that
$(\h\xi)_A = \h$ and $(\h\xi)_B = \xi$. For $\h\in\O$ and $x\in\bZ^d$
we denote by $\th_x \h$ the configuration $\h$ shifted by $x$, namely
$(\th_x\h)_y := \h_{y-x}$, $y\in\bZ^d$. 
Given a probability measure $\mu$ and a random variable $f$ we denote
by $\mu (f)$ the expectation of $f$ w.r.t.\ $\mu$ and by $\mu (f;f):=
\mu \big( f - (f) \big)^2$ its variance.

For $\eta\in\O$ we let $T_{x,y} \eta \equiv\eta^{x,y}$ be
the configuration obtained from $\eta$ by exchanging the number of
particles in $x$ and $y$, i.e.\
\be{1}
(T_{x,y} \eta)_z :=
\left\{
\begin{array}{ll}
\eta_y & \textrm{ if \ } z=x\\
\eta_x & \textrm{ if \ } z=y\\
\eta_z  & \textrm{ if \ } z\neq x,y
\end{array}
\right.
\end{equation}
Analogously, we let $T_{x} \eta \equiv\eta^{x}$ be the
configuration obtained from $\eta$ by flipping the occupation number
in $x$, i.e.\
\be{2}
(T_{x} \eta)_z :=
\left\{
\begin{array}{ll}
1-\eta_x & \textrm{ if \ } z=x\\
\eta_z  & \textrm{ if \ } z\neq x
\end{array}
\right.
\end{equation}
We let also $T_x$, resp.\ $T_{x,y}$, act on functions $f:\O \to \bR$ as 
$T_{x}f (\eta) := f (T_{x}\eta)$, resp. $T_{x,y}f (\eta) := f
(T_{x,y}\eta)$. We finally introduce $\nabla_x f := T_xf -f$ and
$\nabla_{x,y}f := T_{x,y} f - f$.  

The models we consider are defined as follows. 
Given a positive integer $\ell$, let $\L:= [1,\ell]^d
\cap \bZ^d$, $\L^\complement=\bZ^d\setminus\L$.
We consider the continuous time Markov process on the configuration
space $\O_\L$ with generator
\be{3}
L_\L := L_{\textrm{bulk}} + \frac{1}{\ell}  \, L_{\textrm{bound.}}
\end{equation}
where $L_{\textrm{bulk}}$ describes the exchanges of particles in bulk;
it is given by 
\be{4}
L_{\textrm{bulk}} f \, (\eta)= 
\sum_{ \genfrac{}{}{0pt}{1}{\{x,y\}\subset\Lambda }{\dis(x,y)=1} }
c_{x,y}(\eta) \nabla_{x,y} f (\eta)
\end{equation}
namely the occupation numbers at the sites $x,y\in\L$ are exchanged
with rate $c_{x,y}$.
On the other hand
$L_{\textrm{bound.}}$ describes the effect of the particle
reservoirs at the boundary of $\L$; it is given by 
\be{5}
L_{\textrm{bound.}} f \, (\eta)= 
\sum_{
\genfrac{}{}{0pt}{1}{x\in\Lambda, y \not\in\Lambda}{\dis(x,y)=1}
}
c_x(\eta)\nabla_{x} f (\eta)
\end{equation}
For $\rho\in (0,1)$ and $x\in\L$ we choose the flip rates $c_x(\eta)$ as
\be{cg}
c_x(\eta) := (1-\rho) \eta_x +\rho (1-\eta_x)
\end{equation}
Therefore $L_{\textrm{bound.}}$ is the generator of the following
process. A particle on the interior boundary of $\L$ leaves the system
with rate $1-\rho$ (in fact at higher rate at the corners) while
particles enter the system, if the landing site is empty, with rate
$\rho$.
We emphasize that in \eqref{3} the boundary part of the dynamics is
slowed by a factor $1/\ell$, as we prove below this is the minimal 
choice to obtain that the relaxation time diverges as $\ell^2$.

We let $0\in\O$ be the configuration in which all the sites are empty.
%Given $\eta\in\O_\L$, we let $\tilde\eta\in\O$ be given by
%$\tilde \eta := \eta 0_{\L^\complement}$ for $x\in\L$ and $\eta_x=0$
%for $x\not\in\L$. 
We shall discuss two specific choices of the exchange rates
$c_{x,y}(\h)$, $\h\in\O_\L$.
The first is 
\be{6} c^{(1)}_{x,x+e_i} (\eta) := 
\left\{
\begin{array}{ll}
0 & \textrm{ if \ }  (\eta 0_{\L^\complement})_{x-e_i}+
(\eta 0_{\L^\complement})_{x+2e_i} =2, ~~~ i=1,\dots,d
\\
1 & \textrm{ otherwise \ } 
\end{array}
\right.
\end{equation}
namely the exchange across the bond $\{x,x+e_i\}$ is suppressed if the
neighboring sites in the $i$ direction are both occupied.
Note that, since $(\eta 0_{\L^\complement})_x=0$ for $x\not\in\L$,
exchanges across the bonds $\{x,x+e_i\}$ such that either
$x-e_i\in\L^\complement$ or $x+2e_i\in\L^\complement$
are not suppressed.

The second is
\be{7}
c^{(2)}_{x,y} (\eta) :=
\left\{
\begin{array}{ll}
0 & \textrm{ if \ }  {\displaystyle 
\vphantom{\Big\{}
\sum_{z : \: \dis (\{z\},\{x,y\}) =1} 
(\eta 0_{\L^\complement})_z > 2d-1}
\\
1 & \textrm{ otherwise \ } 
\end{array}
\right.
\end{equation}
namely the exchange across the bond $\{x,y\}$ is suppressed if more
than one half of the neighboring sites are occupied.  Note that for
$d=1$ we have ${c}^{(1)}_{x,y}={c}^{(2)}_{x,y}$.  We shall denote by
$L_\L^{(1)}$, respectively $L_\L^{(2)}$ the generator \eqref{3} with
$c_x$ chosen as in \eqref{cg} and $c_{x,y}=c_{x,y}^{(1)}$,
respectively $c_{x,y}=c_{x,y}^{(2)}$.

Let $\mu_{\L,\rho}$ be the Bernoulli measure on $\O_\L$ with density
$\rho$, i.e.\ $\mu_{\L,\rho}$ is the product measure on $\O_\L$ with
marginal $\mu_{\L,\rho}(\eta_x=1)=\rho$. It is easy to check that the
generator $L_\L^{(k)}$, $k=1,2$, is self--adjoint in
$L_2(\mu_{\L,\rho})$, equivalently the rates satisfy detailed balance
w.r.t.\ $\mu_{\L,\rho}$.

\smallskip
We note that the bulk dynamics $L_{\textrm{bulk}}$ preserves the total
number of particles in $\L$, but - since the rates degenerate - it is
not ergodic on all the hyperplanes of $\O_\L$ with fixed total number
of particles. For istance, if $d=1$, all configurations $\eta$ in
which the distance between all the empty sites is three or more do
not evolve. On the other hand, thanks to $L_{\textrm{bound.}}$, it is
not difficult to show that the generator $L_\L^{(k)}$, $k=1,2$, is
irreducible, namely there is positive probability of going from any
configuration to any other. By standard theory on finite state space
Markov chain, irreducibility of $L^{(k)}_\L$ implies the uniqueness of the
invariant measure and that $0$ is a simple eigenvalue of $L^{(k)}_\L$.
In Section \ref{s:g} we prove a lower bound on the spectral gap of
$L^{(k)}_\L$ in $L_2(\mu_{\L,\rho})$ showing that for each
$\rho\in(0,1)$ it shrinks to $0$ as $\ell^{-2}$.

\bigskip
In order to discuss the diffusive behavior of the tagged particle we
need to introduce also the infinite volume dynamics. The configuration
space is then $\O =\{0,1\}^{\bZ^d}$, a function $f:\O\longrightarrow \bR$
is called a {\em local\/} function if it depends only on finitely
many $\h_x$. The generator of the process acts on local functions as 
\be{9}
\cL^{(k)} f \, (\eta)= 
\sum_{ \genfrac{}{}{0pt}{1}{\{x,y\}\subset\bZ^d}{\dis(x,y)=1} }
c_{x,y}^{(k)}(\eta) \nabla_{x,y} f (\eta)
\end{equation}
where $c^{(k)}$, $k=1,2$ has been defined in \eqref{6},\eqref{7}, where
now we have $\L^\complement=\es$. Note that $c^{(k)}$ are
translationally covariant in the sense that 
$c^{(k)}_{x+y,x+y+e}(\th_y \h)= c^{(k)}_{x,x+e}(\h)$. Moreover, for
each $\r\in [0,1]$ and$k=1,2$, the generator 
$\cL^{(k)}$  is self adjoint in $L_2(\mu_\r)$, where
$\mu_\r$ is the Bernoulli measure in $\O$ with density $\r$.
In Section \ref{s:tp} we prove that, for each $\r\in[0,1]$ and $k=1,2$, the
generator $\cL^{(k)}$ is ergodic in $L_2(\mu_\r)$, namely
that $0$ is a simple eigenvalue.

We consider the process $\h(t)$ generated by $\cL^{(k)}$ and condition
that at time zero the origin is occupied; we then tag the particle at
the origin and denote by $x(t)$ its position at time $t$. 
The pair $\big(\h(t),x(t)\big)$ is then a Markov process on the state space
$\big\{ (\h,x)\in \O\times \bZ^d  \,:\: \h_x=1\big\}$ 
with generator
\be{10}
\begin{array}{lcl}
{\displaystyle
\cA^{(k)} F \, (\h,x) 
}
&:=&
{\displaystyle
\sum_{ \genfrac{}{}{0pt}{1}{y\in\bZ^d}{\dis(x,y)=1} }
c_{x,y}^{(k)}(\eta) (1-\h_y)\big[ F(\eta^{x,y},y) -F(\h,x) \big]
}\\
&&{\displaystyle
+\sum_{ \genfrac{}{}{0pt}{1}{\{y,z\}\subset\bZ^d\sm\{x\}}{\dis(y,z)=1} }
c_{y,z}^{(k)}(\eta) \big[ F(\eta^{y,z},x) -F(\h,x) \big]
}
\end{array}
\end{equation}

Let $\O_0:=\{\h\in\O\,\: : \h_0=1\}$ and $\mu_{\r,0}$ be the Bernoulli
measure on $\O_0$ with marginal $\mu_{\r,0}(\h_x=1)=\r$,
$x\in\bZ^d\sm\{0\}$. We shall consider the process $\big(\h(t),x(t)\big)$
generated by $\cA^{(k)}$ with initial condition $x(0)=0$ and $\h(0)$
distributed according to $\mu_{\r,0}$. In Section \ref{s:tp}, for
$d\ge 2$, we  prove the invariance principle for the position of the tagged
particle, namely that $\e x(\e^{-2}t)$ converges in distribution, as
$\e\to 0$, to a Brownian motion with strictly positive diffusion
coefficient.

\sezione{Spectral gap}{s:g}
\par\noindent
The spectral gap of the Markov generator $L_\L$ is defined as
$$
\gap(L_\L) := \inf {\rm spec}\, (- L_\L \restriction \id^\perp  )
$$
where $\id^\perp$ is the subspace of $L_2(\mu_{\L,\rho})$ orthogonal
to the constant functions. Since $\O_\L$ is finite, by irreducibility
of $L_\L$, we trivially have $\gap(L_\L)>0$, we next discuss its
asymptotic behavior as $\ell\to\infty$.

In the case of non--degenerate rates $c_{x,y}=1$, namely for the
symmetric simple exclusion process, the aforementioned problem of non
ergodicity on hyperplanes is not present. 
Let $\nu_{\L,n} (\eta):= \mu_{\L,\rho} \big( \eta \big|
\sum_{x\in\L}\eta_x=n \big)$ be the canonical measure with $n$
particles.  In \cite[\S 8]{Q} it is proven that, considering
$L_{\textrm{bulk}}$ with $c_{x,y}=1$ on $L_2(\nu_{\L,n})$, we have $\gap(
L_{\textrm{bulk}}) \asymp \ell^{-2}$ uniformly in $n$, 
here $a_\ell \asymp b_\ell$
means there exists a constant $C>0$ such that $C^{-1} b_\ell \le
a_\ell \le C b_\ell$ for any $\ell>0$. 
In the case $c_{x,y}=1$ it is not difficult to prove, and in fact it
is a corollary of our analysis, that also
for $L_\L$ on $L_2(\mu_{\L,\rho})$ we have  
$\gap( L_\L) \asymp \ell^{-2}$ uniformly in $\r$.

Our first results is a lower bound on the spectral gap of
$L_\L^{(k)}$, $k=1,2$. 
Let us define the Dirichlet form associated to $L_\L^{(k)}$ as 
\be{8}
\begin{array}{lcl}
{\displaystyle \cE_{\L,\rho}^{(k)}(f) } &:=& {\displaystyle  
- \mu_{\L,\rho}( f L_\L^{(k)} f)
\vphantom{\Bigg\{} 
}
\\
&=& {\displaystyle  \frac{1}{2} \Bigg\{ 
\sum_{ \genfrac{}{}{0pt}{1}{\{x,y\}\subset\Lambda }{\dis(x,y)=1} }
\mu_{\L,\rho} \big[ c_{x,y}^{(k)} (\nabla_{x,y} f)^2 \big]
+\frac{1}{\ell} 
\sum_{
\genfrac{}{}{0pt}{1}{x\in\Lambda, y \not\in\Lambda}{\dis(x,y)=1}
}
\mu_{\L,\rho}\big[ c_{x} (\nabla_{x} f)^2 \big]
\Bigg\}
}
\end{array}
\end{equation}

\bteo{t:gap}
For any
$\rho\in(0,1)$ and $k=1,2$ there exists a constant $C=C(d,\rho,k)$ such that
for any $\ell$ and for any function 
$f:\O_\L\rightarrow\bR$ we have

\be{gap}
\mu_{\L,\rho}(f;f) \le C \, \ell^2 \, \cE_{\L,\rho}^{(k)}(f)
\end{equation}

\eteo

\noindent{\it Remark 1.\ } 
Thanks to the variational characterization of the spectral gap, the
bound \eqref{gap} is equivalent to $\gap(L_\L^{(k)}) \ge C^{-1}
\ell^{-2}$. Moreover, by letting $P_t^{(k)}:= \exp\{t L_\L^{(k)}\}$ be
the semigroup generated by $L_\L^{(k)}$, we also have that \eqref{gap}
is equivalent to
$$
\big\| P_t^{(k)} f -\mu_{\L,\rho} f \big\|_{L_2(\mu_{\L,\rho})} 
\le \exp\Big\{ - \frac{t}{C \ell^2} \Big\} \: 
\big\| f -\mu_{\L,\rho} f \big\|_{L_2(\mu_{\L,\rho})}
$$
for any function $f$ on $\O_\L$.
Finally, let $\bE_{\mu_{\L,\rho}}^{(k)}$ be the distribution of 
the stationary process
generated by $L_\L^{(k)}$, i.e.\  we take $\mu_{\L,\rho}$ as the starting
measure of the Markov chain. Then \eqref{gap} is equivalent to
$$
\bE_{\mu_{\L,\rho}}^{(k)} \big( f(\eta(0)) ;  f(\eta(t)) \big)
\le \exp\Big\{ - \frac{t}{C \ell^2} \Big\} \: 
\big\| f -\mu_{\L,\rho} f \big\|_{L_2(\mu_{\L,\rho})}^2  
$$
for any function $f$ on $\O_\L$.

\smallskip
\noindent{\it Remark 2.\ } 
By taking as test function $f(\eta)=\sum_{x\in\L} (\eta_x -\rho) \cos
\frac{\pi x}{2\ell} $ and using $c_{x,y}^{(k)}\le 1$ for any $x,y$ and $k=1,2$, a simple computation
shows that for each $\rho\in (0,1)$ there exist a constant
$C=C(d,k,\rho)$ such that $\gap(L_\L^{(k)}) \le C \ell^{-2}$. Hence
$\gap(L_\L^{(k)})\asymp \ell^{-2}$ as in the case of the simple
exclusion.

\smallskip
\noindent{\it Remark 3.\ } 
As discussed in the introduction, 
the correct dependence of the spectral gap on the density $\rho$ has
some interest.  For simplicity we discuss it only in the
case $k=1$, namely for the rates chosen as in \eqref{6}. It is a
corollary of our analysis that the gap goes to zero as $\rho\uparrow 1$ as
a power law of exponent between 1 and 2.  
More precisely the following two inequalities hold. There exists a
constant $C_1=C_1(d)$ such that for any $\rho\in(0,1)$ and any $\ell$ 
we have $\gap(L^{(1)}_\L) \ge (1-\rho)^2 C_1/\ell^2 $. 
For each integer $\ell\ge 5 $, there exists a constant $C_2=C_2(d,\ell)$ such
that $\gap(L^{(1)}_\L)\le C_2 (1-\rho)$.  
The lower bound follows from the proof of Theorem \ref{t:gap}; 
the upper bound is obtained easily by using as test function
$f(\h)= \h_x$ with $x\in\L$ such that $\dis(x,\L^\complement) \ge 3$.

\smallskip
\noindent{\it Remark 4.\ } 
As previously discussed, the process generated by
$L_{\textrm{bulk}}$ is not irreducible on the hyperplanes with fixed
number of particles, $\O_{\L,N} :=\big\{\h\in\O_\L \,:\:
\sum_{x\in\L}\h_x =N \big\}$. In the one--dimensional case, $d=1$ 
(recall that in such a case $c^{(1)}=c^{(2)}$),
is however not difficult to check that $L_{\textrm{bulk}}$ is irreducible
on the set
$$
\widetilde\O_{\L,N} :=\big\{\h\in\O_{\L,N} \,:\: \exists \, x,y\in\L,\,
x\neq y\,, \dis(x,y)\le 2\, \textrm{ such that } \h_x=\h_y=0 \big\}
$$
and $L_{\textrm{bulk}}$ satisfies detailed balance w.r.t.\ the
conditional measure 
$\widetilde\nu_{\L,N}(\cdot) := \mu_{\L,\r}(\cdot| \widetilde\O_{\L,N})$.
A natural question is then the asymptotic behavior of the spectral
gap of $L_{\textrm{bulk}}$ in $L_2(\widetilde\nu_{\L,N})$, a
reasonable guess is that for each $N\le \ell -2$ we still have
$\gap(L_{\textrm{bulk}})\asymp \ell^{-2}$. This conjecture is supported by
the fact that if $N=|\L|-2=\ell-2$, i.e.\ in the highest density case,
we have a single pair of neighboring empty sites which performs a
random walk.
We are not able to prove the above conjecture in general, but only in
the trivial situation in which $N < |\L|/3=\ell/3$. In such a case,
for $\ell$ large enough, we have $\widetilde\O_{\L,N}= \O_{\L,N}$;
therefore the statement follows easily by a comparison with the
exclusion process with long exchanges, see \cite[Lemma 8.1]{Q}, and a
minor modification of the argument in Lemma \ref{t:cam} below.

\bigskip
The key step in the proof of Theorem \ref{t:gap} is the following
lemma, which is in the same spirit of the path lemmata in 
\cite{CM,CMR,LY,Q,Y}.

\blem{t:cam}
For $k=1,2$ and each $x\in\L$ let 
$$
U_x^{(k)} := 
\big\{ y\in \L \,: \: 1 \le y_1 \le x_1 - 1 \,,\: |y_i-x_i|\le r^{(k)},
\,i=2,\dots,d \big\}
$$
where $r^{(1)}:=0$ and $r^{(2)}:=1$. 
Then, for each $k=1,2$ and $\rho\in(0,1)$, there exists a constant 
$A=A(d,k,\rho)$ such that for any $\ell$
\be{cam}
\mu_{\L,\rho} \big( c_x (\nabla_x f)^2 \big)
\le A \Big\{ \ell \, \sum_{y\in U_x^{(k)}} 
\mu_{\L,\rho}\big( c_{y,y+e_1}^{(k)} (\nabla_{y,y+e_1} f)^2 \big)
+ \sum_{ \newatop{y:\, y_1=1}{y\in U_x^{(k)}}}
\mu_{\L,\rho}\big( c_{y} (\nabla_{y} f)^2 \big)
\Big\}
\end{equation}
for any $x\in\L$ and any function $f$ on $\O_\L$.
\elem

Postponing the proof of the lemma above, let us first show how it
implies, together with a comparison argument with Glauber dynamics,
Theorem \ref{t:gap}.

\smallskip
\noi{\it Proof of Theorem \ref{t:gap}.} \
Let us introduce the product (Glauber) dynamics in $\O_\L$ 
defined by the generator
\be{g}
L^G_\L f (\eta) := \sum_{x\in\L} c_x(\eta) \nabla_x f (\eta)
\end{equation}
where $c_x$ has been defined in \eqref{cg}. The generator $L^G_\L$ is
self--adjoint in $L_2(\mu_{\L,\rho})$; since it is a product dynamics, it
is immediate to check its spectral gap is $1$. For each function $f$ on $\O_\L$
we thus get
\be{1.0}
\begin{array}{l}
{\displaystyle\mu_{\L,\rho}(f;f) \le 
- \mu_{\L,\rho} (f L^G_\L f) =  
\frac 12 \sum_{x\in\L} \mu_{\L,\rho} \big( c_x (\nabla_x f)^2 \big)
}
\\
{\displaystyle
\phantom{\mu_{\L,\rho} (f;f)}
\le 
\frac{A}{2} \sum_{x\in\L} \Big\{ \ell \, \sum_{y\in U_x^{(k)}} 
\mu_{\L,\rho} \big( c_{y,y+e_1}^{(k)} (\nabla_{y,y+e_1} f)^2 \big)
+ \sum_{ \newatop{y:\, y_1=1}{y\in U_x^{(k)}}} 
\mu_{\L,\rho} \big( c_{y} (\nabla_{y} f)^2 \big)
\Big\}
}
\\
{\displaystyle
\phantom{\mu_{\L,\rho} (f;f)} \le
\frac{A}{2} \ell (2 r^{(k)}+1)^{d-1}
\Big\{ \ell \, \sum_{ 
\genfrac{}{}{0pt}{1}{\{x,y\}\subset\Lambda }{\dis(x,y)=1}
}
\mu_{\L,\rho} \big( c_{x,y}^{(k)} (\nabla_{x,y} f)^2 \big)
+
\sum_{
\genfrac{}{}{0pt}{1}{x\in\Lambda, y \not\in\Lambda}{\dis(x,y)=1}
}
\mu_{\L,\rho} \big( c_x (\nabla_{x} f)^2 \big)
\Big\}
}
\end{array}
\end{equation}
where we used the variational characterization of the spectral gap, Lemma \ref{t:cam} and elementary inequalities.
We thus get the bound \eqref{gap} with $C=A(2r^{(k)}+1)^{d-1}$.
\qed

\medskip
We are left with the proof of the lemma. The basic idea is to use 
first $L_{\textrm{bound.}}$ to empty a few sites at the boundary. 
Then - via careful moves - we show how this cluster of holes can be shifted,
using exchanges with non zero rate, and used to flip the occupation number in
$x$. Finally we shift the cluster back to the boundary and use 
again $L_{\textrm{bound.}}$ to reconstruct the initial configuration 
near the boundary.

\smallskip
\noi{\it Proof of Lemma \ref{t:cam}.} \
We discuss first the case of $k=1$ which corresponds to the rates
\eqref{6}. We assume also that $x_1\ge 4$, otherwise the
proof is much easier.

Given $\eta\in\O_\L$ and $x\in\L$ let us define $S_x\eta\in\O_\L$ as the
configuration given by
\be{sxe}
(S_x\eta)_y :=
\left\{
\begin{array}{ll}
0 & \textrm{ if \ } y=(1,x_2,\dots,x_d) \\
0 & \textrm{ if \ } y=(2,x_2,\dots,x_d) \\
1-\eta_x & \textrm{ if \ } y=(3,x_2,\dots,x_d) \\
\eta_y  & \textrm{ otherwise \ }
\end{array}
\right.
\end{equation}
Moreover, for $y\in\bZ^d$ we define
\be{tlr}
\begin{array}{lcl}
\TL_y &:=& T_{y+e_1,y+2e_1}T_{y,y+e_1}T_{y-e_1,y} \\
\TR_y &:=& T_{y,y+e_1}T_{y+e_1,y+2e_1}T_{y+2e_1,y+3e_1} 
\end{array}
\end{equation}
Note that $\TL_y$ moves the occupation number in $y-e_1$ to
$y+2e_1$ while the configuration in $y,y+e_1,y+2e_1$ is shifted by one
in the direction $-e_1$. 
Analogously $\TR_y$ moves the occupation number in $y+3 e_1$ to
$y$ while the configuration in $y,y+e_1,y+2e_1$ is shifted by one
in the direction $e_1$.

For $x\in\L$, we let $\g_i:=(ie_1,x_2,\dots,x_d)$, $i=1,\dots x_1 -1$
and define
\be{bsx}
\ol{S}_x \h := 
\big( \TL_{\g_2}  %\TL_{(3,x_2,\dots,x_d)}
	\cdots \TL_{\g_{x_1-3}} \big)  
T_{x-e_1,x} 
\big( \TR_{\g_{x_1-4}}  %\TR_{(x_1-5,x_2,\dots,x_d)}
	\cdots \TR_{\g_1} \big)  
S_x\h
\end{equation}
It is not difficult to check that 
$(\ol{S}_x \h)_y =0$ if $y=\g_1$ or  $y=\g_2$,
$(\ol{S}_x \h)_y =\eta_x$ if $y=\g_3$,
$(\ol{S}_x \h)_y =1-\eta_x$ if $y=x$,
and $(\ol{S}_x \h)_y =\eta_y$ otherwise.

For $x\in\L$ with $x_1\ge 4$ we write
\be{nf=}
\nabla_x f (\eta) 
=\big[ f(\h^x)- f( \ol{S}_x\h ) \big]  
+ \big[ f( \ol{S}_x\h ) - f(S_x\h) \big]
+ \big[ f({S}_x\h ) - f(\h) \big]
\end{equation}
We start by considering the second term in decomposition above; we claim that
\be{1.1}
%\begin{array}{rcl}
{\displaystyle 
\sum_{\h\in\O_\L} \mu_{\L,\r}(\eta) c_x(\eta) 
\big[ f( \ol{S}_x\h ) - f(S_x\h) \big]^2
}
%&\le & 
{\displaystyle 
\le 18 (1-\r)^{-2} (2x_1 -7) \sum_{ i = 1}^{x_1-1}  
\mu_{\L,\r} \Big[ 
c^{(1)}_{\g_i, \g_{i+1}}
\big( \na_{\g_i, \g_{i+1}} f \big)^2 
\Big]
}%\\
%\phantom{merda}
%&&
%{\displaystyle +
%\sum_{\h\in\O} \mu_\r(\eta) c_x(\eta) \big[ f( {S}_x\h ) - f(\h) \big]^2
%}
%\end{array}
\end{equation}
To prove the above bound, let us introduce the path
\be{1.2}
\z_i:=
\left\{
\begin{array}{ll}
S_x\h
& i=0\\
\TR_{\gamma_i} \cdots \TR_{\gamma_1}   
S_x\h& i=1,\dots,x_1-4
\\
T_{x-e_1,x} \big( \TR_{\gamma_{x_1-4}}  \cdots \TR_{\gamma_1} \big)  
S_x\h & i=x_1-3 
\\
\big( \TL_{\g_{2x_1-5-i}}  %\TL_{(3,x_2,\dots,x_d)}
	\cdots \TL_{\g_{x_1-3}} \big)
T_{x-e_1,x} 
\big( \TR_{\gamma_{x_1-4}}  \cdots \TR_{\gamma_1} \big) 
S_x\h
& i=x_1-3+1,\dots,2 x_1-7 
\end{array}
\right.
\end{equation}
By \eqref{bsx} we have $\z_{2 x_1-7}=\ol{S}_x\h$.  
Note also that, for each $\h\in\O_\L$ and $0\le i\le x_1 -4$, the
configuration $\z_i$ is guaranteed to be empty at the sites $\g_{i+1}$
and $\g_{i+2}$. Moreover, for each $\h\in\O_\L$ and $0 \le j \le x_1 -4$
the configuration $\z_{x_1-3+j}$ is guaranteed to be empty at the
sites $\g_{x_1-2-j}$ and $\g_{x_1-3-j}$.
This will allow us to move the configuration $\z_i$ to the
configuration $\z_{i+1}$ using exchanges with non zero $c^{(1)}$ rate.

By telescopic sums and Cauchy--Schwartz, we get
\be{1.3}
\big[ f( \ol{S}_x\h ) - f(S_x\h) \big]^2
=
\bigg(\sum_{i=1}^{2 x_1-7}  \big[ f( \z_i ) - f(\z_{i-1}) \big] \bigg)^2
\le (2 x_1-7)
\sum_{i=1}^{2 x_1-7} \big[ f( \z_i ) - f(\z_{i-1}) \big]^2
\end{equation}
We consider only the case $1\le i \le x_1-4$, the others are analogous.
We then have $\z_i=\TR_{\g_i}\z_{i-1}$; recalling
\eqref{tlr}, again by telescopic sums and Cauchy--Schwartz, we get
\be{1.4}
\begin{array}{l}
{\displaystyle \vphantom{\bigg\{}
\big[ f( \z_i ) - f(\z_{i-1}) \big]^2 = 
\big[ f( \TR_{\g_i} \z_{i-1} ) - f(\z_{i-1}) \big]^2
}
\\
\quad
{\displaystyle \vphantom{\bigg\{}
\le 3\Big\{ \big[ 
f(T_{\g_i,\g_i+e_1}T_{\g_i+e_1,\g_i+2e_1}T_{\g_i+2e_1,\g_i+3e_1}\z_{i-1}) 
- f(T_{\g_i+e_1,\g_i+2e_1}T_{\g_i+2e_1,\g_i+3e_1}\z_{i-1}) 
\big]^2
}
\\
\quad 
{\displaystyle \phantom{mer}  \vphantom{\bigg\{}
+\big[ 
f(T_{\g_i+e_1,\g_i+2e_1}T_{\g_i+2e_1,\g_i+3e_1}\z_{i-1}) 
- f(T_{\g_i+2e_1,\g_i+3e_1}\z_{i-1}) 
\big]^2
}
\\
\quad 
{\displaystyle \phantom{mer}  \vphantom{\bigg\{}
+\big[ 
f(T_{\g_i+2e_1,\g_i+3e_1}\z_{i-1}) 
- f(\z_{i-1}) 
\big]^2
\Big\}
}
\\
\quad 
{\displaystyle \vphantom{\bigg\{}
= 3\Big\{ 
c^{(1)}_{\g_i,\g_i+e_1}(T_{\g_i+e_1,\g_i+2e_1}T_{\g_i+2e_1,\g_i+3e_1}\z_{i-1}) 
\big[ \nabla_{\g_i,\g_i+e_1}
f(T_{\g_i+e_1,\g_i+2e_1}T_{\g_i+2e_1,\g_i+3e_1}\z_{i-1})  \big]^2
}
\\
\quad 
{\displaystyle \phantom{mer}  \vphantom{\bigg\{}
+c^{(1)}_{\g_i+e_1,\g_i+2e_1}(T_{\g_i+2e_1,\g_i+3e_1}\z_{i-1}) 
\big[ \nabla_{\g_i+e_1,\g_i+2e_1}
f(T_{\g_i+2e_1,\g_i+3e_1}\z_{i-1})  \big]^2
}
\\
\quad 
{\displaystyle \phantom{mer}  \vphantom{\bigg\{}
+c^{(1)}_{\g_i+2e_1,\g_i+3e_1}(\z_{i-1}) 
\big[ \nabla_{\g_i+2e_1,\g_i+3e_1} f(\z_{i-1})  \big]^2
\Big\}
}
\end{array}
\end{equation}
where we used that 
$$
\begin{array}{lcl}
\displaystyle {c^{(1)}_{\g_i+2e_1,\g_i+3e_1}(\z_{i-1})
}&=&
\displaystyle {
c^{(1)}_{\g_i+e_1,\g_i+2e_1}(T_{\g_i+2e_1,\g_i+3e_1}\z_{i-1})
 }\\
&&\vphantom{\}}\\
&=& {\displaystyle
c^{(1)}_{\g_i,\g_i+e_1}(T_{\g_i+e_1,\g_i+2e_1}T_{\g_i+2e_1,\g_i+3e_1}
\z_{i-1}) 
 }=1
\end{array}
$$ 
by construction of the path $\z_i$, see the remark below \eqref{1.2}.

In order to prove the bound \eqref{1.1} we consider only the last term on
the r.h.s.\ of \eqref{1.4}, the other can be analyzed in the same way.
Given $A\su\L$, $\xi\in\O_A$, $x\in A$, and $g:\O_\L\rightarrow \bR$ 
we observe that, since $\mu_{\L,\r}$ is a product measure and $c_x$ is
given in \eqref{cg},
\be{ac}
\begin{array}{lcl}
{\displaystyle
\sum_{\h\in\O_\L} \mu_{\L,\r}(\h) \, \h_x \, c_x(\h) \: 
g(\h_{\L\sm A} \, \xi) 
}
&=&
{\displaystyle
\sum_{\h\in\O_\L} \mu_{\L,\r}(\h) \, (1-\h_x) \, c_x(\h) \: 
g(\h_{\L\sm A} \, \xi) 
}
\\
&=& 
{\displaystyle\vphantom{\Big\{}
\r(1-\r) \mu_{\L,\r} \big( g \,\big|\, \h_A=\xi \big)
}
\end{array}
\end{equation}
Recalling definitions \eqref{sxe} and \eqref{1.2},
%by using that the measure $\mu_{\L\sm\{x,\g_i,\g_{1+1},\g_{i+2}\},\r}$ 
%is invariant w.r.t.\ $\TR_{\g_j}$, $1\le j\le i-1$ 
we thus get
\be{1.8}
\begin{array}{l}
{\displaystyle 
\sum_{\h\in\O} \mu_{\L,\r}(\eta) c_x(\eta) 
c^{(1)}_{\g_i+2e_1,\g_i+3e_1}(\z_{i-1}) 
\big[ \nabla_{\g_i+2e_1,\g_i+3e_1} f(\z_{i-1})  \big]^2
}
\\
\qquad
{\displaystyle = 
\sum_{\h\in\O} \mu_{\L,\r}(\eta) \eta_x c_x(\eta) 
c^{(1)}_{\g_i+2e_1,\g_i+3e_1}(\z_{i-1}) 
\big[ \nabla_{\g_i+2e_1,\g_i+3e_1} f(\z_{i-1})  \big]^2
}
\\
\qquad
{\displaystyle \phantom{=} \: +\: 
\sum_{\h\in\O} \mu_{\L,\r}(\eta) [1-\eta_x] c_x(\eta) 
c^{(1)}_{\g_i+2e_1,\g_i+3e_1}(\z_{i-1}) 
\big[ \nabla_{\g_i+2e_1,\g_i+3e_1} f(\z_{i-1})  \big]^2
}
\\
\qquad
{\displaystyle 
= \r(1-\r) 
\bigg\{
\mu_{\L,\r} \Big( c^{(1)}_{\g_i+2e_1,\g_i+3e_1} 
\big[ \nabla_{\g_i+2e_1,\g_i+3e_1} f  \big]^2 
 \,\Big|\, \h_{x} = 1 , \h_{\g_i}=\h_{\g_{i+1}}=\h_{\g_{i+2}}=0 \Big)
}
\\
\qquad
{\displaystyle \phantom{ \r(1-\r) \bigg\{}
+\mu_{\L,\r} \Big( c^{(1)}_{\g_i+2e_1,\g_i+3e_1} 
\big[ \nabla_{\g_i+2e_1,\g_i+3e_1} f  \big]^2 
\,\Big|\, \h_{x}=\h_{\g_i}=\h_{\g_{i+1}}=0,  \h_{\g_{i+2}}=1 \Big)
\bigg\}
}
\end{array}
\end{equation}

We now observe that for any positive function $g: \O_\L\rightarrow \bR$ we have
\be{positive}
\begin{array}{l}
{\displaystyle 
\vphantom{\bigg\{}
\mu_{\L,\rho} (g) 
\ge 
\mu_{\L,\rho} ( \h_{x}=1, \h_{\g_i}=\h_{\g_{i+1}}=\h_{\g_{i+2}}=0)
\, \mu_{\L,\rho} 
\big( g \,\big|\, \h_{x}=1, \h_{\g_i}=\h_{\g_{i+1}}=\h_{\g_{i+2}}=0 \big)
}
\\
{\displaystyle 
\phantom{\mu_{\L,\rho} (g) \ge} \vphantom{\bigg\{}
+\:
\mu_{\L,\rho} ( \h_{x}=\h_{\g_i}=\h_{\g_{i+1}}=0,  \h_{\g_{i+2}}=1)
\,
\mu_{\L,\rho}
\big( g \,\big|\, \h_{x}=\h_{\g_i}=\h_{\g_{i+1}}=0,  \h_{\g_{i+2}}=1 \big)
}
\\
{\displaystyle 
\phantom{\mu_{\L,\rho} (g)} \vphantom{\bigg\{}
= 
\r (1-\r)^3 \Big\{ 
\mu_{\L,\rho} \big( g \,\big|\, \h_{x}=1,
\h_{\g_i}=\h_{\g_{i+1}}=\h_{\g_{i+2}}=0 \big)
}
\\
{\displaystyle 
\phantom{\mu_{\L,\rho} (g) = } \vphantom{\bigg\{}
\phantom{\r (1-\r)^3 \Big\{ }
+
\mu_{\L,\rho}\big( g \, \big|\, \h_{x}=\h_{\g_i}=\h_{\g_{i+1}}=0,
\h_{\g_{i+2}}=1 \big) 
\Big\}
}
\end{array}
\end{equation}
so that from \eqref{1.8} we get
\be{1.10}
\begin{array}{l}
{\displaystyle 
\sum_{\h\in\O_\L} \mu_{\L,\rho} (\eta) c_x(\eta) 
c^{(1)}_{\g_i+2e_1,\g_i+3e_1}(\z_{i-1}) 
\big[ \nabla_{\g_i+2e_1,\g_i+3e_1} f(\z_{i-1})  \big]^2
}
\\
\qquad\qquad
{\displaystyle 
\le (1-\r)^{-2} \, \mu_{\L,\rho}  \Big( 
c^{(1)}_{\g_i+2e_1,\g_i+3e_1}
\big[ \nabla_{\g_i+2e_1,\g_i+3e_1} f  \big]^2
\Big)
}
\end{array}
\end{equation}
The bound \eqref{1.1} follows from \eqref{1.3}, \eqref{1.4}, and
\eqref{1.10}. Note that the extra factor $2$ comes from the return
path, $x_1-4\le i \le 2x_1-7$.

\medskip

Let us now consider the last term on the r.h.s.\ of \eqref{nf=}; we claim that
\be{1.7}
\begin{array}{l}
{\displaystyle 
\sum_{\h\in\O}
\mu_{\L,\rho} c_x(\h) \big[ f(S_x\h)-f(\h) \big]^2
}
\\
\qquad{\displaystyle 
\le
\frac{6}{(1-\rho)^2} \Big\{ 3 \mu_{\L,\rho} \big( c_{\g_1} [\nabla_{\g_1}f]^2\big)
+
2 \mu_{\L,\rho} \big( c_{\g_1,\g_2} [\nabla_{\g_1,\g_2}f]^2\big)
+\mu_{\L,\rho} \big( c_{\g_2,\g_3} [\nabla_{\g_2,\g_3}f]^2\big)
\Big\}
}
\end{array}
\end{equation}

To prove the above bound let us define $T^+_x \h$ as the configuration
given by $(T^+_x\h)_x = 1$ and $(T^+_x\h)_y = \h_y$ for $y\neq x$;
analogously we let $T^-_x\h$ be the configuration given by
$(T^-_x\h)_x = 0$ and $(T^-_x\h)_y = \h_y$ for $y\neq x$.  Recalling \eqref{sxe} we then
have 
\be{tele} 
\begin{array}{l}
{\displaystyle 
\vphantom{\Big\}}
f(S_x\h)-f(\h) =
\h_x \Big[ 
\h_{\g_1} \nabla_{\g_1} f
(\h) + \nabla_{\g_1,\g_2} f (T^-_{\g_1} \h) +\h_{\g_2} \nabla_{\g_1} f
( T_{\g_1,\g_2} T^-_{\g_1} \h)
}
\\
{\displaystyle
\vphantom{\Big\}}
\phantom{f(S_x\h)-f(\h) =\h_x \big[}
+\nabla_{\g_2,\g_3} f ( T^-_{\g_1} T_{\g_1,\g_2} T^-_{\g_1} \h) + 
\nabla_{\g_1,\g_2} f (T_{\g_2,\g_3}
T^-_{\g_1} T_{\g_1,\g_2} T^-_{\g_1} \h) 
}\\
{\displaystyle 
\vphantom{\Big\}}
\phantom{f(S_x\h)-f(\h) =\h_x \big[}
+ \h_{\g_3} \nabla_{\g_1} f(T_{\g_1,\g_2} T_{\g_2,\g_3} T^-_{\g_1} T_{\g_1,\g_2} T^-_{\g_1} \h)
\Big]
}\\
{\displaystyle 
\vphantom{\bigg\}}
\phantom{f(S_x\h)-f(\h)=}
+(1- \h_x)\Big[ (1-\h_{\g_1}) \nabla_{\g_1} f (\h) 
+\nabla_{\g_1,\g_2} f (T^+_{\g_1} \h) +\h_{\g_2} \nabla_{\g_1} f (
T_{\g_1,\g_2} T^+_{\g_1} \h) 
}
\\
{\displaystyle 
\vphantom{\Big\}}
\phantom{f(S_x\h)-f(\h)=}
\phantom{+ (1-\h_x)\big[}
+ \nabla_{\g_2,\g_3} f ( T^-_{\g_1}
T_{\g_1,\g_2} T^+_{\g_1} \h) 
+\nabla_{\g_1,\g_2} f (T_{\g_2,\g_3}
T^-_{\g_1} T_{\g_1,\g_2} T^+_{\g_1} \h) 
}
\\
{\displaystyle 
\vphantom{\Big\}}
\phantom{f(S_x\h)-f(\h)=}
\phantom{= + (1-\h_x)\big[}
+ \h_{\g_3} \nabla_{\g_1} f
(T_{\g_1,\g_2} T_{\g_2,\g_3} T^-_{\g_1} T_{\g_1,\g_2} T^+_{\g_1} \h)
\Big]
}
\end{array}
\end{equation}

By using Schwartz inequality and the fact that the 
telescopic decomposition in \eqref{tele} has been arranged so that all the exchanges 
have non zero $c^{(1)}$ rate, we get 
\begin{equation}
\begin{array}{l}
{\displaystyle \vphantom{\Big\{}
[f(S_x\h)-f(\h)]^2
\leq
6\Big\{ 
\h_x~
\h_{\g_1} [\nabla_{\g_1} f (\h)]^2
+ (1- \h_x) (1-\h_{\g_1})[\nabla_{\g_1} f (\h)]^2
}
\\
{\displaystyle \vphantom{\Big\{}
\phantom{\le 6\Big\{ }
+ \h_xc^{(1)}_{\g_1,\g_2}(T^-_{\g_1} \h) 
[\nabla_{\g_1,\g_2} f (T^-_{\g_1} \h)]^2 
+ (1- \h_x)c^{(1)}_{\g_1,\g_2}(T^+_{\g_1} \h) 
[\nabla_{\g_1,\g_2} f (T^+_{\g_1} \h)]^2
}
\\
{\displaystyle \vphantom{\Big\{}
\phantom{\le 6\Big\{ }
+ \h_x \h_{\g_2} 
 [ \nabla_{\g_1} f ( T_{\g_1,\g_2} T^-_{\g_1} \h)]^2 
+ (1- \h_x) \h_{\g_2} 
[\nabla_{\g_1} f ( T_{\g_1,\g_2} T^+_{\g_1} \h)]^2 
}
\\
{\displaystyle \vphantom{\Big\{}
\phantom{\le 6\Big\{ }
+ \h_x c_{\g_2,\g_3}^{(1)} (T^-_{\g_1} T_{\g_1,\g_2} T^-_{\g_1} \h) 
[\nabla_{\g_2,\g_3} f ( T^-_{\g_1} T_{\g_1,\g_2} T^-_{\g_1} \h)]^2
}
\\
{\displaystyle \vphantom{\Big\{}
\phantom{\le 6\Big\{ }
+ (1- \h_x)  c_{\g_2,\g_3}^{(1)}( T^-_{\g_1} T_{\g_1,\g_2} T^+_{\g_1} \h)
[\nabla_{\g_2,\g_3} f ( T^-_{\g_1} T_{\g_1,\g_2} T^+_{\g_1} \h)]^2 
}\\
{\displaystyle \vphantom{\Big\{}
\phantom{\le 6\Big\{ }
+\h_x 
c_{\g_1,\g_2}^{(1)}(T_{\g_2,\g_3} T^-_{\g_1} T_{\g_1,\g_2} T^-_{\g_1} \h)
[\nabla_{\g_1,\g_2} 
f (T_{\g_2,\g_3} T^-_{\g_1} T_{\g_1,\g_2} T^-_{\g_1} \h)]^2 
}\\
{\displaystyle \vphantom{\Big\{}
\phantom{\le 6\Big\{ }
+(1- \h_x) c_{\g_1,\g_2}^{(1)} (T_{\g_2,\g_3} T^-_{\g_1} T_{\g_1,\g_2} T^+_{\g_1} \h)
[\nabla_{\g_1,\g_2} f (T_{\g_2,\g_3} T^-_{\g_1}
T_{\g_1,\g_2} T^+_{\g_1} \h)]^2
}\\
{\displaystyle \vphantom{\Big\{}
\phantom{\le 6\Big\{ }
+ \h_x \h_{\g_3} [\nabla_{\g_1} f
(T_{\g_1,\g_2} T_{\g_2,\g_3} T^-_{\g_1} T_{\g_1,\g_2} T^-_{\g_1}
\h)]^2
}
\\
{\displaystyle \vphantom{\Big\{}
\phantom{\le 6\Big\{ }
+ (1- \h_x) \h_{\g_3} [\nabla_{\g_1} (f (T_{\g_1,\g_2}
T_{\g_2,\g_3} T^-_{\g_1} T_{\g_1,\g_2} T^+_{\g_1} \h)]^2 
\Big\}
}
\end{array}
\label{tele2}
\end{equation}

By recalling the definition \eqref{cg} we have 
$1-\h_y\leq \r^{-1} c_y(\h)$ and 
$\h_y \leq (1-\r)^{-1} c_y(\h)$ for any $\h\in\O_\L$ and $y\in\L$.
We next estimate separately each term on the r.h.s.\ of \eqref{tele2}.
Let us consider only the last two terms, the others
are easier. We have
\be{1.7.5}
\begin{array}{l}
{\displaystyle 
\sum_{\h\in\O_\L}
\mu_{\L,\rho} (\h)~ c_x(\h) 
\Big\{  \h_x~ \h_{\g_3} [\nabla_{\g_1} f
(T_{\g_1,\g_2} T_{\g_2,\g_3} T^-_{\g_1} T_{\g_1,\g_2} T^-_{\g_1}
\h)]^2
}
\\
{\displaystyle 
\phantom{\sum_{\h\in\O}\mu_{\L,\rho} (\h) c_x(\h) \Big\{ }
+ (1- \h_x)~ \h_{\g_3} [\nabla_{\g_1} (f (T_{\g_1,\g_2}
T_{\g_2,\g_3} T^-_{\g_1} T_{\g_1,\g_2} T^+_{\g_1} \h)]^2 
\Big\}
}
\\
\qquad
{\displaystyle 
\le (1-\r)^{-1} 
\Big\{ (1-\r)~ \mu_{\L,\rho} (\h_x) 
\mu_{\L,\rho}\big( c_{\g_1} [ \nabla_{\g_1} f]^2 
\,\big|\, \h_x=1,\h_{\g_2}=\h_{\g_3}=0\big)
}
\\
\qquad
{\displaystyle 
\phantom{\le (1-\r)^{-1}  \Big\{}
+ \r~ \mu_{\L,\rho} (1-\h_x) 
\mu_{\L,\rho} \big( c_{\g_1} [ \nabla_{\g_1} f]^2 
\,\big|\, \h_x=\h_{\g_2}=0, \h_{\g_3}=1\big)
\Big\}
}
\\
\qquad
{\displaystyle \vphantom{\Big\{}
\le  (1-\r)^{-2} \mu_{\L,\rho} \big(  c_{\g_1} [ \nabla_{\g_1} f]^2  \big)
}
\end{array}
\end{equation}
where we used that, as in \eqref{positive}, for any positive function
$g:\O_\L\rightarrow \bR$ 
$$
\mu_{\L,\rho} \big( g \,\big|\, \h_x=1,\h_{\g_2}=\h_{\g_3}=0\big)
+ \mu_{\L,\rho} \big( g \,\big|\, \h_x=\h_{\g_2}=0, \h_{\g_3}=1\big)
\le \frac{1}{\r(1-\r)^2}
\, \mu_{\L,\rho}\big( g \big)
$$
By analogous computations for the other terms, \eqref{1.7} follows.

\medskip
Finally, to bound the first term on the r.h.s.\ of \eqref{nf=}, it is
enough to change variable $\h\mapsto\h^x$. Indeed, noting $\bar{S}_x\h^x=S_x\h$, we get
\be{1.20}
\sum_{\h\in\O_\L}
\mu_{\L,\rho} (\h) c_x(\h) \big[ f(\h^x)-f(\bar S_x\h) \big]^2
=
\sum_{\h\in\O_\L}
\mu_{\L,\rho} (\h) c_x(\h) \big[ f(\h)-f(S_x\h) \big]^2
\end{equation}

For $x$ such that $x_1\ge 4$, the bound \eqref{cam}, with the 
constant $A$ given by  $A=180 (1-\r)^{-2}$ now follows from 
\eqref{nf=}, \eqref{1.1}, \eqref{1.7}, and \eqref{1.20}.
The case in which $1\le x_1\le 3$ can be proven directly by the same
steps leading to \eqref{1.7}. 

\bigskip

\begin{figure}[h]
\centerline{    \epsfysize=8cm
       \epsffile{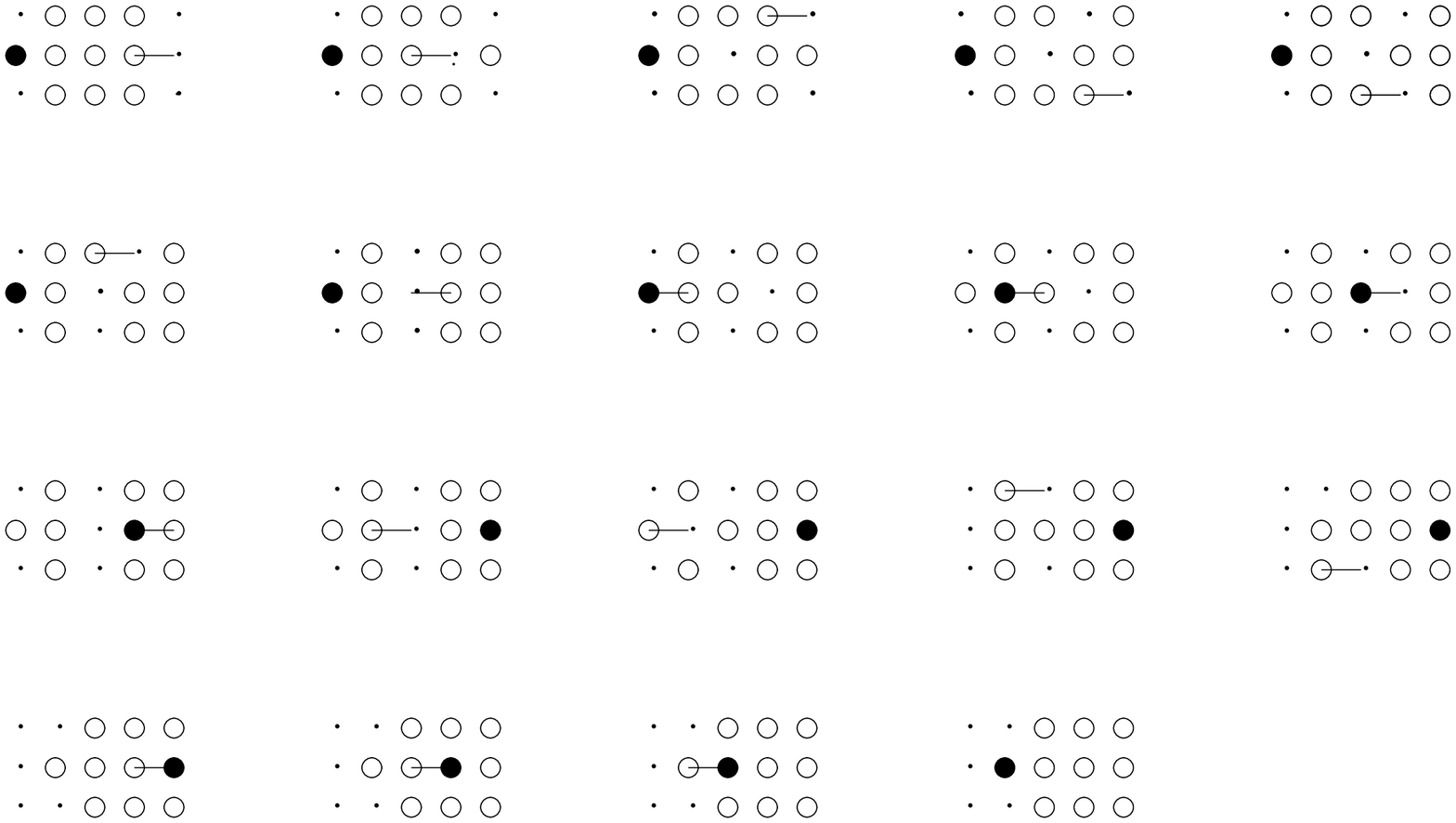}}
\caption{{\it The shifting path.\/} $\circ$ denotes sites guaranteed empty,
$\bullet$ denotes $1-\h_x$, $\cdot$ denotes an arbitrary occupation
number, and --- denotes the bond exchanged in the
next move.}
\label{f:path}
\end{figure}

In the case $k=2$, namely for the choice \eqref{7} of the exchange
rates, we give only a rough sketch of the proof, which is very similar
to the case $k=1$. Indeed, it is enough to define the
configuration $S_x\h$, analogous to \eqref{sxe}, as
\be{sxe2}
(S_x\eta)_y :=
\left\{
\begin{array}{ll}
0 & \textrm{ if \ } y\in Q_{1}((3,x_2,\dots,x_d)) \\
1-\eta_x & \textrm{ if \ } y=(1,x_2,\dots,x_d) \\
\eta_y  & \textrm{ otherwise \ }
\end{array}
\right.
\end{equation}
where $Q_{1}(x):=\{y\in \bZ^d \,: \: \max_{i=1,\dots,d} |x_i-y_i|\le
1\}$ is the cube of side $3$ centered in $x$. 
The configuration $S_x\eta$ can then be shifted using exchanges with
non zero $c^{(2)}$ rates by means of a suitable path,
depicted in Fig.\ \ref{f:path} for $d=2$.
The proof is finally completed by the same arguments given for the
choice \eqref{6}.\qed

\sezione{Log--Sobolev inequality and entropy decrease 
for the porous media equation}{s:ls}

\par\noindent
In this Section, we show how the techniques introduced in the proof
of the spectral gap can be used to prove a logarithmic Sobolev
inequality for exclusion processes with degenerate rates. 
Moreover, by a scaling limit of the logarithmic Sobolev inequality, we
deduce the exponential decrease of a suitable entropy for a
nonlinear degenerate parabolic equation.

We recall that a Markov process with reversible measure $\mu$ and
Dirichlet form $\cE$ is said to satisfy a logarithmic Sobolev
inequality with constant $c_{\textrm{{\tiny LS}}}$ iff for any function $f$ 
we have
\be{ls0}
\mu\bigg( f^2 \log \frac{ f^2}{\mu( f^2)} \bigg) \le
c_{\textrm{{\tiny LS}}} \, \cE(f)
\end{equation}
It is well known, see e.g.\ \cite{cyril}, that \eqref{ls0} implies the
hypercontractivity of the semigroup $P_t$ associated to the Markov
process and the exponential decay of the entropy, namely, for any
probability density $f$ w.r.t.\ $\mu$ we have
$$
\mu\Big( P_t f \log  P_t f \Big) \le 
\exp\Big\{ - \frac{4 \, t}{c_{\textrm{{\tiny LS}}}}\Big\}
\: \mu\big( f \log   f \big)
$$

We first state the logarithmic Sobolev inequalities for the process
generated by $L^{(k)}_\L$, $k=1,2$. Recall the Dirichlet form 
$\cE_{\L,\rho}^{(k)}$ has been defined in \eqref{8}.

\bteo{t:ls}
For each $k=1,2$ and $\rho\in(0,1)$ there exists a constant
$C=C(d,k,\rho)$ such that for any $\ell$ and for any function 
$f:\O_\L\rightarrow\bR$ we have
\be{ls}
\mu_{\L,\rho} \bigg(f^2 \log \frac{ f^2}{\mu_{\L,\rho}( f^2)} \bigg)
\le C \, \ell^2 \, \cE_{\L,\rho}^{(k)}(f)
\end{equation}
\eteo

\noi{\it Proof of Theorem \ref{t:ls}.} \
Let $L_\L^G$ be the generator on $\O_\L$ introduced in \eqref{g}; then,
see e.g.\ \cite{cyril}, for any $\rho\in(0,1)$ it satisfies the logarithmic
Sobolev inequality \eqref{ls0} with 
$c_{\textrm{\tiny LS}}$ given by $C_1(\rho):= (1-2\r)^{-1}\,\log[(1-\r)/\r]$
(we understand $C_1(1/2)=2$) uniformly in $\ell$.  
By \eqref{cam} and \eqref{1.0} the bound \eqref{ls}, with $C=
C_1(\rho) \, A\, (2 r^{(k)}+1)^{d-1}$ (recall $A=A(d,\rho,k)$ is the
constant in Lemma \ref{t:cam}), follows. \qed

\bigskip
Let us consider the following parabolic problem, called porous media
equation, on $B:=[0,1]^d$ with Dirichlet boundary conditions 
\be{pme}
\left\{
\begin{array}{ll}
\partial_t u (t,r) = \nabla_r \cdot\big( D(u (t,r) ) \nabla_r u (t,r) \big) 
& (t,r)\in(0,\infty)\times B  \\
u(t,r)=\rho 
&(t,r)\in(0,\infty)\times \partial B
\\
u(0,r)= \f(r) 
& r\in B
\end{array}
\right.
\end{equation}
where $\rho\in (0,1)$, the initial datum $\f\in
C(B;[0,1])$ satisfies $\f(r)=\rho$ for $r\in\partial B$, 
and the diffusion coefficient $D(u)\ge 0$ is
smooth and degenerates linearly for $u=1$, namely the exists a
constant $\d\in(0,1)$ such that $\d (1-u)\le D(u) \le \d^{-1} (1-u)$,
$u\in [0,1]$. Since we assumed $0\le\f\le 1$, by the maximum
principle, we have that $u\in C(\bR_+\times B; [0,1])$.

As discussed in the introduction, the equation \eqref{pme} is the
natural candidate for the hydrodynamic limit of the process with
generator $L_\L^{(k)}$, the diffusion coefficient $D(u)$ would be
given by a Green--Kubo formula \cite[\S II.2.2]{S}.  Note that the
Dirichlet boundary condition is due to the particles' reservoirs.
Although we do not prove any scaling limit of the microscopic dynamics
to \eqref{pme}, we show how the logarithmic Sobolev inequality
\eqref{ls} implies the exponential decrease of a suitable entropy for
the nonlinear evolution \eqref{pme}.

Given $\rho\in(0,1)$, we introduce the convex 
functional $H_\rho : C(B; [0,1]) \rightarrow  \bR_+$ as
\be{dhr}
H_\r(u) := \int_{B}\!dr \:
\Big[ u(r)\log\frac{u(r)}{\r} + (1-u(r)) \log\frac{1-u(r)}{1-\r} \Big]
\end{equation}
where we understand $0\log 0 =0$.
It is easy to show that $H_\rho$ is a Lyapunov functional for the
evolution \eqref{pme}, moreover if $u(t,r)$ is a smooth solution of
\eqref{pme} bounded away from 0 and 1 we have
\be{dhdt}
- \frac{d}{dt} H_\r(u(t,\cdot)) =
\int_{B} \!dr \: u(t,r)[1-u(t,r)] D(u(t,r)) 
\Big( \nabla_r \log \frac{u(t,r)}{1-u(t,r)}\Big)^2
=:\cQ(u(t,\cdot))
\end{equation}

The following theorem, which states a  ``logarithmic Sobolev inequality'' 
for the nonlinear evolution \eqref{pme} is easily obtained as a
scaling limit of \eqref{ls}.

\bteo{t:lspe}
For each $\rho\in(0,1)$ and $\d>0$ there exists a constant
$C'=C'(d,\d,\rho)$ such that for any $u\in C^1 \big(B; [0,1] \big)$ 
with $u(r)=\r$ for $r\in \partial B$ 
\be{lspe}
H_\r(u) \le C' \cQ(u)
\end{equation}
\eteo

\noindent{\it Remark.\ } 
The inequality \eqref{lspe} can be proven directly by reducing it to
the Poincar\'e inequality for the Dirichlet Laplacian on $B$.
The probabilistic proof given below, which somehow connects the
evolution \eqref{pme} to the microscopic process, shows additionally
that the Lyapunov functional $H_\r$ is the macroscopic limit of a
relative entropy.

\medskip
\noi{\it Proof.} \
We shall prove the bound \eqref{lspe} for $D(u)=1-u^2$; 
the generic case of $D$ degenerating linearly for $u\uparrow 1$ 
follows by our hypothesis on the diffusion coefficient $D$.
By truncation, it is also enough to prove \eqref{lspe} when $u$ is a smooth
function bounded away from 0 and 1. 

We set $\e:=\ell^{-1}$ and apply inequality \eqref{ls} for
$k=1$ choosing $f^2=g_\e$ where
\be{dge}
g_\e(\h) = \prod_{x\in\L} \frac{ u(\e x)^{\h_x} [1-u(\e x)]^{1-\h_x}}
{ \r^{\h_x} [1-\r]^{1-\h_x}}
\end{equation}
Note that $\mu_{\L,\r,u}^\e(\h):= \mu_{\L,\r}(\h) g_\e(\h)$ 
is a product probability measure on $\O_\L$ with density profile $u$,
namely $\mu_{\L,\r,u}^\e(\h_x) = u(\e x)$. By elementary
computations which we omit, we have that the normalized relative
entropy of $\mu_{\L,\r,u}^\e$ w.r.t.\ $\mu_{\L,\r}$ converges to
$H_\r(u)$ as $\e\to 0$, namely
\be{er}
%\lim_{\e\to 0} \e^d \cH \big( \mu_{\L,\r,u}^\e \big| \mu_{\L,\r}
%\big) := 
\lim_{\e\to 0} \e^d 
\mu_{\L,\rho} \Big(g_\e \log \frac{ g_\e }{\mu_{\L,\rho}(g_\e)} \Big)
= H_\r(u)
\end{equation}
Moreover it is straightforward to check that
\be{lls}
\lim_{\e\to 0} \e^{d-2} 
\cE_{\L,\rho}^{(1)}(\sqrt{g_\e})
= \cQ(u)
\end{equation}
Let $C(d,1,\r)$ be the constant such that \eqref{ls} holds for $k=1$.
By \eqref{er} and \eqref{lls} the bound \eqref{lspe}, with 
$C'=C(d,1,\r)$, now follows from Theorem \ref{t:ls}. 
\qed

\medskip
The exponential decrease of the ``entropy'' $H_\r$ along the flow
of the porous media equation \eqref{pme} follows from \eqref{dhdt},
Theorem \ref{t:lspe}, and a straightforward truncation argument.

\bcor{t:dee}
Let $u\in C(\bR_+\times B; [0,1])$ be the solution of
\eqref{pme} and $C'=C'(d,\d,\rho)$ be the constant in \eqref{lspe}.
For each $\rho\in(0,1)$ we have
\be{dee}
H_\r( u(t,\cdot) ) \le e^{ - t/C' } H_\r(\f )
\end{equation}
for any  $t\in\bR_+$ and any 
$\f\in C(B; [0,1])$ such that $\f(r)=\rho$ for $r\in\partial B$.
\ecor

We have assumed that the diffusion coefficient $D(u)$ degenerates
linearly for $u= 1$. One can also obtain the exponential decrease
of the entropy $H_\r$ if $D(u) \asymp (1-u)^n$, $n$ a positive
integer. This can be shown by introducing a microscopic model in which
the exchange rate $c_{x,x+e_i}(\h)$ is zero iff there exists
$j=1,\dots,n$ such that $\h_{x-je_i}=\h_{x+(j+1)e_i}=1$. It is
in fact possible to prove that the logarithmic Sobolev constant 
for such a model is of the order $\ell^2$.

\sezione{Diffusion of the tagged particle}{s:tp}
\par\noindent
In this Section we consider stochastic lattice gases with degenerate
rates in infinite volume. We first prove that, for each $\r\in[0,1]$, 
the generator $\cL^{(k)}$, $k=1,2$ defined in \eqref{9} is ergodic in 
$L_2(\mu_\r)$, recall that $\mu_\r$ is the Bernoulli measure on $\O$.

\bpro{t:ea0}
For each $\r\in [0,1]$ and $k=1,2$ we have that 
zero is a simple eigenvalue of the generator 
$\cL^{(k)}$ considered on $L_2(\mu_{\r})$.  
\epro

\noi{\it Proof.} \ 
Let
\be{dfl}
\cE^{(k)}_\r (f) =
\frac{1}{2} 
\sum_{ \genfrac{}{}{0pt}{1}{\{x,y\}\subset\bZ^d }{\dis(x,y)=1} }
\mu_{\rho} \big[ c_{x,y}^{(k)} (\nabla_{x,y} f)^2 \big]
\end{equation}
be the Dirichlet form of the generator $\cL^{(k)}$. To 
show zero is a simple eigenvalue of $\cL^{(k)}$ 
we check that  $\cE^{(k)}_\r(f)=0$ implies $f$ constant $\mu_{\r}$--a.s.
This is trivially true for $\r=0,1$. 
For $\r\in(0,1)$, by De Finetti's theorem, it is enough to show 
that $\cE^{(k)}_\r(f)=0$ implies $\mu_{\r}(\nabla_{x,y}f)^2 =0$ for each $\{x,y\}\su\bZ^d$
with $\dis(x,y)=1$.

We discuss in some detail the case $k=1$.
Let $x\in\bZ^d$ and consider the bond 
$\{x,x+e_i\}$, $i=1,\dots,d$. For $n=1,2,\dots$ we introduce the events 
$$
B_{x,i}^{n} :=\big\{\h\in\O \,:\:
\h_{x+ne_i}=\h_{x+(n+1)e_i}=0\big\}
\quad\quad\quad
B_{x,i}:=\bigcup_{n\ge 1} B_{x,i}^{n}
$$
By noting  that $\mu_{\r}(B_{x,i})=1$, we have 
\be{4.2}
\mu_{\r}(\nabla_{x,x+e_i}f)^2
=\mu_{\r}\big( [\nabla_{x,x+e_i}f]^2 \id_{B_{x,i}} \big)
\le \sum_{n=1}^\infty
\mu_{\r}\big( [\nabla_{x,x+e_i}f]^2 \id_{B_{x,i}^{n}} \big)
\end{equation}

Let $\g_h:=x+he_i$, $h=0,1,\dots$; given $\h\in B_{x,i}^{n}$ we can
find a path $\h=\z_0,\dots,\z_N=\h^{x,x+e_i}$ where
$\z_{j+1}=\z_{j}^{\g_h,\g_{h+1}}$ for some $h=0,1,\dots$ and
$c_{\g_h,\g_{h+1}}^{(1)}(\z_{j})=1$.  It is in fact possible to
construct a path analogous to the one introduced in the proof of Lemma
\ref{t:cam}; note that the two sites $\g_n$ and $\g_{n+1}$ are empty
by the definition of the event $B_{x,i}^{n}$.
Since $\cE^{(1)}_\r (f)=0$ implies
$$
\mu_{\r} \big( c_{ \g_h,\g_{h+1}}^{(1)} [\nabla_{\g_h,\g_{h+1}}f]^2 \big)=0
\quad \textrm{for any } h=0,1,\dots 
$$
by telescopic sums and Cauchy--Schwartz in \eqref{4.2} 
we get $\mu_{\r}(\nabla_{x,x+e_i}f)^2=0$. 

Recalling Figure \ref{f:path} it is straightforward 
to modify the argument given above to cover also the case $k=2$. 
\qed

\bigskip
We next discuss the diffusive behavior of a tagged particle. More
precisely, we consider the process $\big(\h(t),x(t)\big)$ with
generator given in \eqref{10}, initial condition $x(0)=0$ and
$\h(0)$ distributed according to $\mu_{\r,0}$, the Bernoulli measure
on $\O_0=\{\h\in\O\,:\:\h_0=1\}$. 
Let $\xi(t):= \th_{-x(t)} \h(t)$ be the process as seen from the tagged
particle, we have that $\xi(t)$ is itself a Markov process on the
configuration space $\O_0$ with generator given by
\be{4.1}
\cA_0^{(k)} f \, (\xi) =
\sum_{ \genfrac{}{}{0pt}{1}{y \in\bZ^d}{\dis(0,y)=1} }
c_{0,y}^{(k)}(\xi) (1-\xi_y)\big[ f(\th_{-y} \xi^{0,y}) -f(\xi) \big]
+\sum_{ \genfrac{}{}{0pt}{1}{\{x,y\}\subset\bZ^d\sm\{0\}}{\dis(x,y)=1} }
c_{x,y}^{(k)}(\xi) \nabla_{x,y} f(\xi) 
\end{equation}

A straightforward computation shows that $\cA_0^{(k)}$ is
self--adjoint in $L_2(\mu_{\r,0})$; moreover, by the same argument 
as in Proposition \ref{t:ea0}, it is also ergodic in
$L_2(\mu_{\r,0})$.  
We can therefore apply the same proof as the one given in \cite{KV,S} 
for non--degenerate rates.
We get that the rescaled position of the tagged particle, $\e x(\e^{-2}t)$, 
%for the process with generator \eqref{10} 
converges in distribution, as $\e\to 0$, to a $d$--dimensional
Brownian motion with diffusion matrix $2\,D^{(k)}_{\textrm{self}}$. 
Furthermore the diffusion matrix
$D^{(k)}_{\textrm{self}}=D^{(k)}_{\textrm{self}}(\r)$ is given by the
variational formula
\be{vfd}
\begin{array}{lcl}
{\displaystyle
r \cdot D^{(k)}_{\textrm{self}}(\r) r 
}
&=&
{\displaystyle
\frac 12 \inf_{f {\textrm{ local}}} \int\! \mu_{\r,0}(d\xi) \: \Big\{
\sum_{ \genfrac{}{}{0pt}{1}{y \in\bZ^d}{\dis(0,y)=1} }
c_{0,y}^{(k)}(\xi) (1-\xi_y)
\big[ r\cdot y + f(\th_{-y} \xi^{0,y})-f(\xi)\big]^2
}\\
&&{\displaystyle
\phantom{\frac 12 \inf_{f {\textrm{local}}} \int\!d\mu_{\r,0}(\xi) \: \Big\{}
+\sum_{ \genfrac{}{}{0pt}{1}{\{x,y\}\subset\bZ^d\sm\{0\}}{\dis(x,y)=1} }
c_{x,y}^{(k)}(\xi) [\nabla_{x,y} f(\xi)]^2 \Big\}
}
\end{array}
\end{equation}
where $r\in\bR^d$ and $\cdot$ is the inner product in $\bR^d$. 

The main result of this Section is that, for $d\ge 2$ and each
$\r\in [0,1)$, the diffusion matrix $D^{(k)}_{\textrm{self}}(\r)$ 
is strictly positive as in the case of simple exclusion \cite{KV,S}.

\bteo{t:dnd}
For each $d\ge 2$, $k=1,2$ and $\r\in[0,1)$ there exists a real
$c=c(d,k,\r) >0$
such that $r \cdot D^{(k)}_{\normalfont\textrm{self}}(\r) r \ge c \, r\cdot r$ 
for any $r\in\bR^d$.
\eteo

\noindent{\it Remark.\ } 
As discussed in the introduction,
the behavior of $D^{(k)}_{\textrm{self}}(\r)$ as $\r\uparrow 1$ 
has some interest. 
Note that for SEP it vanishes linearly. In the case $k=1$, 
Theorem \ref{t:dnd} will be proven with $c = c_0 (1-\r)^{11}$ where
$c_0$ does not depend on $\rho$. 
However, by the same strategy and some extra efforts, it is
possible to improve the lower bound to $c = c_0' (1-\r)^{4}$. An
upper bound of the form $D^{(k)}_{\textrm{self}}(\r) \le C_0 (1-\r)^2
\id$ is easily obtained by using a constant test function $f$ in
\eqref{vfd}.

\smallskip
Let us fix a direction in $\bR^d$, say $e_1$ and define the
following subsets of $\bZ^d\sm\{0\}$
\be{4.3}
\begin{array}{lcl}
{\displaystyle
R_0^{(1)}
}
&:=&
{\displaystyle
\big\{x\in \bZ^d\sm\{0\} \,:\: \max_{i=1,2} |x_i| = 1\,,
\:x_i=0\,,i=3,\dots,d\big\}
}\\
{\displaystyle
R_{\pm 1}^{(1)} 
}
&:=&
{\displaystyle
\big\{x\in \bZ^d\sm\{0\} \,:\: x_1=\pm 2 \,,\: |x_2|\le 1\,,\:
x_i=0\,,i=3,\dots,d\big\}
}
\end{array}
\end{equation}
and
\be{4.4}
\begin{array}{lcl}
{\displaystyle
R_0^{(2)}
}
&:=&
{\displaystyle
\big\{x\in \bZ^d\sm\{0\} \,:\: x_1=0\,,\: 
\max_{i=2,\dots,d} |x_i| \le 3 
\big\}
}\\
{\displaystyle
R_{\pm 1}^{(2)} 
}
&:=&
{\displaystyle
\big\{x\in \bZ^d\sm\{0\} \,:\: x_1=\pm 1 \,,\: 
\max_{i=2,\dots,d} |x_i| \le 3 
\big\}
}
\end{array}
\end{equation}
Given $\xi\in\O_0$, we next define $\xi^{+,-,(k)}$ as the
configuration obtained from $\xi$ by exchanging the occupation numbers
in $R^{(k)}_{+1}$ with the corresponding ones in $R^{(k)}_{-1}$,
namely
\be{4.5}
\big(\xi^{+,-,(1)}  \big)_x :=
\left\{
\begin{array}{ll}
\xi_x & \textrm{ if \ } x\not\in R^{(1)}_{+1}\cup R^{(1)}_{-1}
\\
\xi_{x\mp 4e_1} & \textrm{ if \ } x\in R^{(1)}_{\pm 1}
\end{array}
\right.
\end{equation}
and
\be{4.5bis}
\big(\xi^{+,-,(2)}  \big)_x :=
\left\{
\begin{array}{ll}
\xi_x & \textrm{ if \ } x\not\in R^{(2)}_{+1}\cup R^{(2)}_{-1}
\\
\xi_{x\mp 2e_1} & \textrm{ if \ } x\in R^{(2)}_{\pm 1}
\end{array}
\right.
\end{equation}
We finally introduce the events
\be{4.6}
\cB^{(k)}_{\pm} := 
\big\{\xi \in\O_0 \,:\: \xi_{R^{(k)}_0}=0\,,\: \xi_{R^{(k)}_{\pm 1}}=0
\big\} 
\,,\quad\quad\quad
\cB^{(k)}:= \cB^{(k)}_{+}\cup\cB^{(k)}_{-} 
\end{equation}
and note that $\xi\in\cB^{(k)}_{+}$ iff $\xi^{+,-,(k)}\in\cB^{(k)}_{-}$.

\blem{t:cam2}
For each $d\ge 2$, $k=1,2$, and $\r\in [0,1)$ there exists a real 
$a=a(d,k,\r)>0$ such that for any $r\in\bR^d$ we have
\be{4.7}
\begin{array}{l}
{\displaystyle
 r \cdot D^{(k)}_{\normalfont\textrm{self}}(\rho) r
\ge  (r\cdot e_1)^2
\, \frac a2 \: \inf_{f {\normalfont\textrm{ local}}} \: 
\int\! \mu_{\r,0}\big( d\xi \big| \cB^{(k)} \big) \: 
\Big\{
\big[ f(\xi^{+,-,(k)})-f(\xi)\big]^2
}
\\
{\displaystyle
\phantom{e_1\cdot D^{(k)}_{\normalfont\textrm{self}}(\rho) e_1 
\ge  (r\cdot e_1)^2 \frac a2 \inf_{f {\normalfont\textrm{ local}}}
\Big\{}
+\sum_{y=\pm 1} \id_{ \{\xi_{R^{(k)}_y} = 0\}}(\xi)
\big[ y + f(\th_{- y e_1} \xi^{0,ye_1})-f(\xi)\big]^2
\Big\}
}
\end{array}
\end{equation}
\elem

\noi{\it Proof.} \ 
We discuss in some detail the case $k=1$.
We note that if $\xi\in\cB^{(1)}$ and $y=\pm 1$ we have
$$
c_{0,ye_1}^{(1)}(\xi) [1-\xi_{ye_1}]
\ge  \id_{ \{\xi_{R^{(1)}_y} = 0\}}(\xi)
$$
since $\mu_{\r,0}(\cB^{(1)})\ge (1-\r)^{11}$, by the same argument as
in \eqref{positive}, from \eqref{vfd} we then get
\be{4.8}
\begin{array}{l}
{\displaystyle 
r\cdot D^{(1)}_{\textrm{self}}(\rho) r 
\ge  
(r\cdot e_1)^2 \frac 12 \inf_{f {\textrm{ local}}} 
\Big\{
\int\! \mu_{\r,0}( d\xi) \:
\sum_{ \genfrac{}{}{0pt}{1}{\{x,y\}\subset\bZ^d\sm\{0\}}{\dis(x,y)=1} }
c_{x,y}^{(1)}(\xi) [\nabla_{x,y} f(\xi)]^2 
}
\\
{\displaystyle
\quad\quad\quad
+ (1-\r)^{11} 
\int\! \mu_{\r,0}\big( d\xi \big| \cB^{(1)} \big) \:
\sum_{y=\pm 1} \id_{ \{\xi_{R^{(1)}_y} = 0\}}(\xi)
\big[ y + f(\th_{- y e_1} \xi^{0,ye_1})-f(\xi)\big]^2
 \Big\}
}
\end{array}
\end{equation}

\begin{figure}[h]
\centerline{    \epsfysize=8cm
       \epsffile{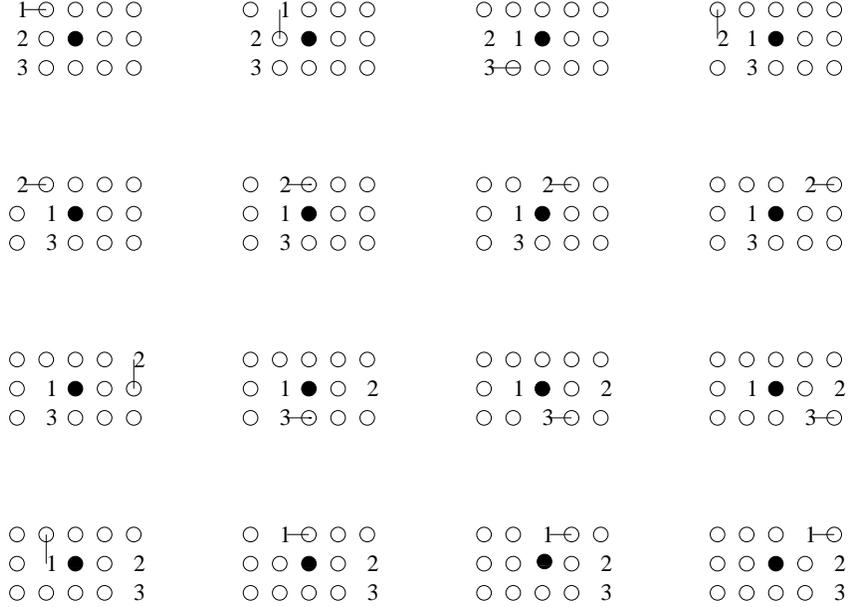}}
\caption{{\it Chain of exchanges $T_1,\dots,T_{16}$ for $k=1$.\/} 
The picture represents sites in the plane $e_1,e_2$,
$\bullet$ denotes site $0$ and --- in the $i$--th figure denotes the bond
exchanged by $T_i$.  
If the sites denoted by $\circ$ are empty then the starting 
configuration is in $\cB^{(1)}_{+}$. In such a case 1,2, and 3 denote
the occupation numbers in $R^{(1)}_-$ which, step by step, are moved to
$R^{(1)}_+$ by using only allowed exchanges.} 
\label{f:path2}
\end{figure}

Let $T_1,\dots,T_{16}$ be the chain of exchanges depicted in Figure
\ref{f:path2}, $T_i$ exchanges the occupation numbers in the bond
$b_i$. Note that if $\xi\in\cB^{(1)}_{+}$ the path 
$\z_0^+:=\xi$, $\z_1^+:=T_1\z_0^+$, $\dots$,
$\z_{16}^+:=T_{16}\z_{15}^+=\xi^{+,-,(1)}$ 
is such that $c_{b_i}^{(1)}(\z_{i-1}^+)=1$, $i=1,\dots, 16$.
For $\xi\in\cB^{(1)}_{-}$ we define analogously 
$\z_0^-:=\xi$, $\z_1^-:=T_{16}\z_0^-$, $\dots$,
$\z_{16}^-:=T_{1}\z_{15}^-=\xi^{+,-,(1)}$ which is such that
$c_{b_{17-i}}^{(1)}(\z_{i-1}^-)=1$, $i=1,\dots, 16$.

We then have
\be{4.9}
\begin{array}{l}
{\displaystyle \vphantom{\Big\{}
\big[f(\xi^{+,-,(1)})-f(\xi)\big]^2 \id_{\cB^{(1)}}(\xi)
}
\\
{\displaystyle \vphantom{\Big\{}
\quad\quad
\le 
\id_{\cB^{(1)}_{+}} (\xi) \big[f(\xi^{+,-,(1)})-f(\xi)\big]^2 
+ \id_{\cB^{(1)}_{-}} (\xi) \big[f(\xi^{+,-,(1)})-f(\xi)\big]^2 
}
\\
{\displaystyle \vphantom{\Big\{}
\quad\quad 
=
\id_{\cB^{(1)}_{+}} (\xi) 
\big[\sum_{i=1}^{16} f(\z_i^+)-f(\z_{i-1}^+)\big]^2 
+ \id_{\cB^{(1)}_{-}} (\xi) 
\big[\sum_{i=1}^{16} f(\z_{i}^-)-f(\z_{i-1}^-)\big]^2 
}\\
{\displaystyle \vphantom{\Big\{}
\quad\quad
\le
16 \sum_{i=1}^{16} \Big\{ 
c_{b_i}^{(1)}(\z_{i-1}^+) \big[ \nabla_{b_i} f (\z_{i-1}^+) \big]^2
+c_{b_{17-i}}^{(1)}(\z_{i-1}^-) \big[ \nabla_{b_{17-i}} f (\z_{i-1}^-) \big]^2
\Big\}
}
\end{array}
\end{equation}
By integrating w.r.t.\ $\mu_{\r,0}$ the above inequality and taking
into account that in the chain of exchanges $T_i$, $i=1,\dots,16$ each
bond is used at most twice we get
\be{4.10}
\begin{array}{l}
{\displaystyle
\mu_{\r,0}\big(\cB^{(1)} \big)
\int\! \mu_{\r,0}\big( d\xi \big| \cB^{(1)} \big) \: 
\big[ f(\xi^{+,-,(1)})-f(\xi)\big]^2
}\\
{\displaystyle \vphantom{\bigg\{}
\quad\quad\quad\quad\quad
= 
\int\! \mu_{\r,0}( d\xi) \: 
\big[f(\xi^{+,-,(1)})-f(\xi)\big]^2 \id_{\cB^{(1)}}(\xi)
}\\
{\displaystyle \vphantom{\Bigg\{}
\quad\quad\quad\quad\quad
\le
64 \int\! \mu_{\r,0}( d\xi) \:
\sum_{ \genfrac{}{}{0pt}{1}{\{x,y\}\subset\bZ^d\sm\{0\}}{\dis(x,y)=1} }
c_{x,y}^{(1)}(\xi) [\nabla_{x,y} f(\xi)]^2 
}
\end{array}
\end{equation}
which inserted in \eqref{4.8} concludes the proof with $a(d,1,\rho)=
2^{-6}\,(1-\r)^{11}$.

\medskip

\begin{figure}[h]
\centerline{    \epsfysize=12cm
       \epsffile{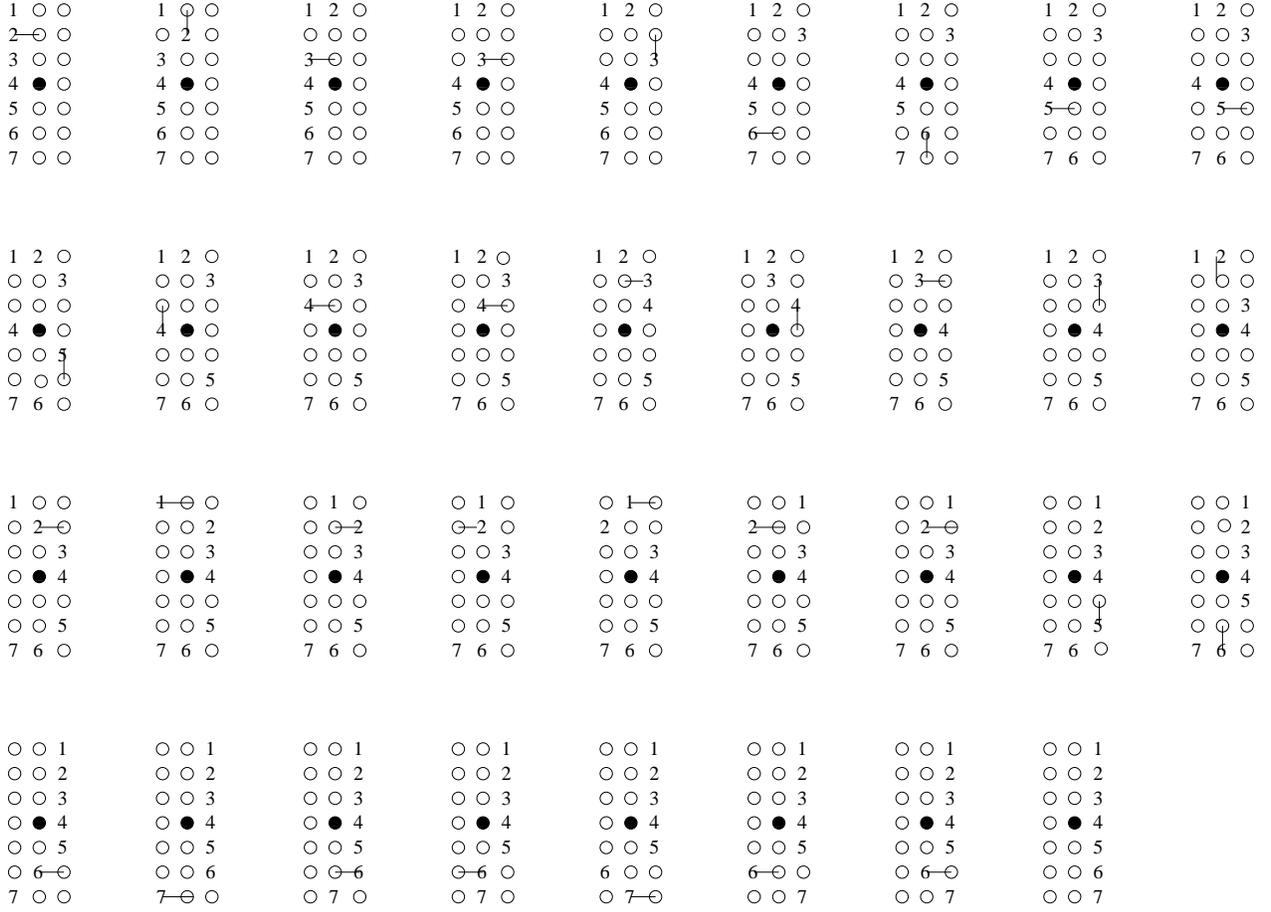}}
\caption{{\it Chain of exchanges $T_1,\dots,T_{35}$ for $k=2$.\/} $\bullet$
denotes site $0$ and --- in the $i$--th figure denotes the bond exchanged
by $T_i$.} 
\label{f:path3}
\end{figure}

The case $k=2$ is proven by the same arguments; in this case, for $d=2$, the 
required chain of exchanges $T_1,\dots,T_{35}$ is depicted 
in Figure \ref{f:path3}.
\qed

\bigskip
\noi{\it Proof of Theorem \ref{t:dnd} (sketch).} \ 
Thanks to Lemma \ref{t:cam2}, it is enough to prove that the right
hand side of \eqref{4.7} is strictly positive for $r\cdot e_1\neq
0$. By the variational formula \eqref{vfd}, it can be interpreted as
the self diffusion coefficient of a one dimensional auxiliary process 
which we next describe in the fixed frame of reference.

The configuration space is $\big\{ (y,\h)\in\bZ\times\O \,:\:
\th_{-ye_1}\h \in \cB^{(k)}\big\}$.  Let $y(t)\in\bZ$ be the position
of the tagged particle and $\h(t)$ be the particles configuration.  At
time $t=0$ the tagged particle is at the origin, $y(0)=0$, and
$\h(0)\in\cB^{(k)}$ is distributed according to $\mu_{\r,0}\big( \cdot
\big| \cB^{(k)} \big)$. Then the tagged particle jumps to the right,
resp.\ left, with rate one if $\th_{-y(t) e_1}\h(t)\in \cB^{(k)}_+$,
resp.\ if $\th_{-y(t) e_1}\h(t)\in \cB^{(k)}_-$. Moreover, with rate
one, $\h(t)$ is exchanged to $\th_{y(t)e_1}\big[ (\th_{-y(t) e_1}
\h(t))^{+,-,(k)} \big]$, namely the occupation numbers in $\th_{y(t)
e_1} R^{(k)}_-$ are exchanged with the ones in $\th_{y(t) e_1}
R^{(k)}_+$.

The proof of the Theorem can now be completed as in the case of
non--degenerate rates, see \cite[II.6.3]{S}, by showing that there
exists 
a real $c>0$ such that for any $t>0$ and $\h\in\cB^{(k)}$ we have
$\bE_{(0,\h)}\big( y(t)^2 \big) \ge c t$. Here $\bE_{(0,\h)}$ denotes
the distribution of the auxiliary process with initial condition $(0,\h)$.
\qed

%%%%% Acknowledgments
\bigskip\noindent
\normalsize\textbf{Acknowledgments}
\par\noindent
{\small 
It is great pleasure to thank G.\ Jona--Lasinio 
for helpful discussions and the constant encouragement. 
We are also grateful to M.\ Bertsch, G.\ Biroli, and F.\ Cesi 
for useful comments. 
}

%%%%%%%% Bibliografia 
%\newpage 
%\addcontentsline{toc}{section}{References} 
%\normalsize\normalfont

\end{document}